\documentclass[12pt]{article}
\usepackage{amsmath,amssymb,amsfonts,amsthm,bm,bbm,cancel,wasysym,mathtools}
\usepackage{epsfig,graphics,graphicx,epstopdf}
\usepackage{array,booktabs,colortbl,colordvi,multirow}
\usepackage{colordvi,color,xcolor}
\usepackage{hyperref}
\usepackage{rotating}

\usepackage{verbatim}
\usepackage{cite}
\usepackage{subfig}
\usepackage{setspace}
\usepackage{url}
\usepackage[percent]{overpic}
\usepackage{slashed}
\usepackage{authblk}
\usepackage{xspace}
\usepackage{fullpage}

\newcommand{\unit}[1]{\ensuremath{\mathrm{\,#1}}\xspace}

\newcommand{\GeV}{\unit{GeV}}

\def\ie{{\it i.e.}}
\def\eg{{\it e.g.}}

\newskip\zatskip \zatskip=0pt plus0pt minus0pt
\def\matth{\mathsurround=0pt}
\def\lsim{\mathrel{\mathpalette\atversim<}}
\def\gsim{\mathrel{\mathpalette\atversim>}}
\def\atversim#1#2{\lower0.7ex\vbox{\baselineskip\zatskip\lineskip\zatskip
  \lineskiplimit 0pt\ialign{$\matth#1\hfil##\hfil$\crcr#2\crcr\sim\crcr}}}

\begin{document}


\begin{flushright}
SLAC-PUB-17155 \\
\today
\end{flushright}
\vspace*{5mm}

\renewcommand{\thefootnote}{\fnsymbol{footnote}}
\setcounter{footnote}{1}

\begin{center}

{\Large {\bf Ghosts- and Tachyon-Free Regions of the Randall-Sundrum }\\
{\bf Model Parameter Space }}\\

\vspace*{0.75cm}
{\bf G. N. Wojcik}~\footnote{gwojci03@stanford.edu}

{\bf J. L. Hewett}~\footnote{hewett@slac.stanford.edu} and 

{\bf T. G. Rizzo}~\footnote{rizzo@slac.stanford.edu},

\vspace{0.5cm}

{SLAC National Accelerator Laboratory, 2575 Sand Hill Rd, Menlo Park, CA, 94025, USA}

\end{center}
\vspace{.5cm}

\begin{abstract}
 
\noindent

Model building within the Randall-Sundrum (RS) framework generally involves placing the Standard Model fields in the bulk. Such fields may possess non-zero values for their associated 
brane-localized kinetic terms (BLKTs) in addition to possible bulk mass parameters. In this paper we clearly identify the regions of the RS model parameter space where the presence of bulk 
mass terms and BLKTs yield a setup which is free from both ghost and tachyon instabilities. Such physically acceptable parameter space regions can then be used to construct realistic and 
phenomenologically viable RS models.  
\end{abstract}

\renewcommand{\thefootnote}{\arabic{footnote}}
\setcounter{footnote}{0}
\thispagestyle{empty}
\vfill
\newpage
\setcounter{page}{1}



\section{Introduction}
\label{Sections/Section_1_Intro}

The Randall-Sundrum (RS) model of warped extra dimensions \cite{Randall:1999ee}, with both the Standard Model (SM) gauge and fermion fields being in the 5-d bulk, provides important insights into two of the most important and outstanding problems that we currently face in particle physics: the Gauge-Hierarchy problem and the Fermion Mass Hierarchy/Flavor Puzzle \cite{eDims,Randall:1999ee,huberHierarchy,gherghetta}.  In order to address these two issues, while also satisfying the numerous experimental constraints arising from collider, flavor and precision measurements \cite{casagrande,huber,carenaEW,Davoudiasl:2000wi,agashe,hewett,rizzoEW,Dey}, requires a highly flexible framework that takes advantage of all the numerous O(1) free parameters that are allowed within the RS model.  Chief among these free parameters are the bulk masses for the various SM fermions that are responsible for the  'localization'  of the fermion wavefunctions within the 5-d bulk, which possess far-reaching consequences for both flavor and neutrino physics \cite{Grossman:2000,hewett,casagrande,huber,Casagrande:2010}. In addition to these bulk mass parameters there are possible localized kinetic terms (BLKTs) \cite{georgi,Davoudiasl:2003zt,Davoudiasl:2002ua,carenaBranes,Aguila:2003,Dey}, on either or both the IR and UV branes, for all of the gauge and fermion SM fields in addition to those that might be present for the graviton. Of course, these various parameters can not be chosen arbitrarily or independently. In addition to the many phenomenological and model-building constraints that are required to be satisfied by any realistic model \cite{casagrande, huber,carenaEW,Davoudiasl:2000wi,agashe,hewett}, one needs to also be concerned about possible unphysical regions of the parameter space wherein ghost and/or tachyon states for the graviton or any of the SM fields may be present in the spectra \cite{Aguila:2003}. Thus the identification of such unphysical regions, $\textit{a priori}$, would be a useful guide in the construction of realistic and phenomenologically successful RS-based models.  Unfortunately, no detailed systematic study of where or when such unphysical regions of the RS model may appear is currently available. The goal of the present paper is to address this situation and provide such a guide.

In order to perform this analysis we first consider the case of a single fermion in the bulk, before electroweak symmetry breaking, with a bulk mass $m=k\nu$ and possessing BLKTs on both the UV(IR) brane described by the parameters $\tau_{0(\pi)}$, respectively.{\footnote {The case of a bulk gauge field or graviton is then analogous to the choice $\nu=-1/2$ or $1/2$, respectively. This is because at these $\nu$ values, the equations of motion for the fermion fields are identical to those of a bulk gauge field (if $\nu=-1/2$) or a graviton (at $\nu=1/2$).}} After determining the general conditions for freedom from both tachyon and ghost instabilities (obtained by considering possible imaginary roots for the eigenvalue equation and the normalization factors of the corresponding eigenfunctions),  for specific values of $\nu$ we determine which values of $\tau_{0,\pi}$ yield equations of motion that result in tachyon- and ghost-free spectra. Specifically, for fixed values of $\nu$, the physically allowed values of $\tau_{0,\pi}$ which lead to either tachyons and/or ghost states are determined. Once this is done, we then investigate the issue of whether or not spontaneous symmetry breaking (SSB) of the SM electroweak symmetry might  influence these results. This requires the consideration of the simultaneous constraints on the two {\it different} fermion fields whose zero-modes we can identify with the specific left- or right-handed SM fermion states. Note that since the SM Higgs vacuum expectation value (vev) ($\sim 246$ GeV) is sufficiently below the phenomenologically allowed Kaluza-Klein (KK) mass scale, $\gsim$a few TeV, we can generally perform this analysis by using a perturbative approach. We then demonstrate that SSB in the perturbative region does not alter our previously results with respect to the physically allowed parameter space regions. 

The outline of this paper is as follows: In Section \ref{Framework} we present a review that provides the necessary background information on the RS model, establishes our essential notation and describes the assumptions to be used in the subsequent analysis. In Section \ref{Setup}, we provide the basic mathematical framework for performing the analysis and describe the procedures that we will subsequently follow. In particular, we divide the relevant  range of the parameter $\nu$ into several distinct regimes that we will discuss separately. We find that this is separation is most easily performed  by considering the shifted parameter $\eta= -(\nu +1/2)$. In Section \ref{TeVFermions}, we consider the range  $\eta \sim -0.1$, which corresponds to a fermion localized close to the IR brane, while in Section \ref{UVFermions} the range $\eta \gsim 0.1$, corresponding to a fermion localized near the UV brane, is instead examined. Note that the latter range includes the case of gravitons, which corresponds to $\eta=1$. The rather complex range $-0.1 \lsim \eta \lsim 0.1$, corresponding to a fermion largely delocalized in the bulk, which includes the case of bulk gauge fields (\ie, $\eta=0$) is considered in detail in Section \ref{GaugeFermions}.  In Section \ref{ComplexMasses}, we look beyond the possibility of purely imaginary tachyonic roots to the case where possible complex roots might exist and determine that if such roots were to exist, they would not correspond to any physical, propagating KK states. In Section \ref{SSBDiscussion} we analyze the possible influence of SM electroweak SSB on our previously obtained results and demonstrate that if SSB can be treated perturbatively these results remain valid and that no new parameter space regions are opened up by SSB.  Our results and conclusions are then summarized in Section \ref{Summary}.


\section{Randall-Sundrum Framework}\label{Framework}

In this section, we provide a brief overview of the incorporation of bulk fermions in a generic RS model framework. The model is constructed on a slice of AdS$_5$ spacetime, with the metric \cite{Randall:1999ee},
\begin{equation}\label{metric}
ds^2 = e^{-2 \sigma}\eta_{\mu \nu}dx^\mu dx^\nu - r_c^2 d\phi^2.
\end{equation}

The fifth dimension, parameterized here by the coordinate $-\pi \leq \phi \leq \pi$, is compactified on an $S^1/Z_2$ orbifold of radius $r_c$, and bounded on both sides by 4-dimensional flat Minkowski branes. Following common naming conventions, we refer to the brane at $\phi=0$ as the UV- or Planck-brane, and the brane at $|\phi|=\pi$ as the IR- or TeV-brane. Here, $\sigma \equiv kr_c |\phi|$, where $k \sim O({\overline M}_{Pl})$ is the curvature scale of the warped space, and $\eta_{\mu \nu}$ is the Minkowski metric in four dimensions. As discussed in \cite{Randall:1999ee}, the gauge-gravity hierarchy may be addressed in this framework if $kr_c \approx 11$, with a natural 4-dimensional Higgs vev being generated at the weak scale while keeping gravity at the Planck scale. For our numerical analyses here we take $kr_c = 11.27$. It has been shown that the size of the extra dimension can be stabilized at approximately this value without fine-tuning of parameters \cite{Goldberger}.

To incorporate fermionic fields in the bulk, we start in the simple scenario where spontaneous symmetry breaking via the Higgs mechanism (and the corresponding mixing of fermion Kaluza-Klein tower states) is neglected. Here, in the case of a bulk fermion field (producing a left-handed chiral SM zero mode fermion), we have the action \cite{Carena:2004zn}
\begin{eqnarray}\label{fermionAction}
 S_F= &\frac{}{} &\int d^4 x \int r_c d\phi \, \sqrt{G} \, 
\left\{V^M_N (\frac{i}{2}\bar\Psi \Gamma^N \partial_M \Psi+h.c.) + [2\tau_0/kr_c ~\, \delta(\phi) \right. \nonumber\\ 
& + & 
\left. 2\tau_\pi/kr_c ~\, \delta(|\phi| - \pi)] \,V^\mu_\nu (i\bar\Psi_L \gamma^\nu \partial_\mu \Psi_L +h.c.) \right. \\
& - &
\left. sgn(\phi) ~m_{f_\Psi} \bar \Psi \Psi  \frac{}{} \right\} \nonumber\,.
\end{eqnarray}
Here, Roman indices denote summation over five dimensions (Greek indices indicate summation over the usual four), while $\sqrt{G}=\sqrt{det(G^{MN})}=e^{-4 \sigma}$, $V^M_\mu=e^\sigma \delta^M_\mu$, $V^4_4=-1$, and $\Gamma^N=(\gamma^\nu,i \gamma_5).$ The bulk mass of $\Psi$ is given by $m_{f_\Psi}= k \nu_f$, where $\nu_f$ is a dimensionless parameter that determines the location of the fermion fields in the bulk. Note that this action includes generic brane-localized kinetic terms (BLKT's) (\ie, represented by $\tau_0$ and $\tau_\pi$), which may arise due to loop effects or as a consequence of a UV completion of the theory for the left-handed, but not right-handed, fermion fields. This is by construction; in order to produce a left-handed chiral zero-mode, the left-handed five-dimensional field is required to be even under the orbifold's $Z_2$ symmetry, while the right-handed fields must be odd. Intuitively, we see that right-handed brane terms will be ineffective here: Since the right-handed fields are $Z_2-$odd, their bulk wave functions vanish at $|\phi|=\pi$ and $\phi=0$, so any additional terms on these branes should not have a significant effect on the physics. Furthermore, as noted in \cite{Carena:2004zn}, if the odd fields lack brane terms at tree-level, they will not be perturbatively generated.

We now introduce the following KK expansion for even (L) and odd (R) fermion fields,
\begin{equation}\label{PsiLR}
\Psi_{L,R}=\sum_n \psi_{L,R}^{(n)}(x)\frac{e^{2\sigma}f_{L,R}^{(n)}(\phi)}{\sqrt{r_c}}.
\end{equation}
Here, $\psi^{(n)}_{L(R)}(x)$ represents the left-(right-)handed 4-dimensional wave function for the $n^{th}$ mode of the Kaluza-Klein (KK) tower, while $f_{L(R)}^{(n)}(\phi)$ represents this field's wave function in the five-dimensional bulk. The mass of the $n^{th}$ KK mode is then denoted by $m_n$. Our goal, as is standard in Kaluza-Klein treatments of extra dimensions, is to achieve an effective 4-dimensional theory with an action of the form
\begin{equation}\label{GoalActionNonSSB}
    S_{4} = \sum_{n} \int d^4 x \left[ {\overline {\psi^{(n)}}}i \slashed{\partial} \psi^{(n)}-m_n {\overline {\psi^{(n)}}} \psi^{(n)} \right].
\end{equation}
To achieve canonically normalized kinetic terms, we require the following normalization condition
\begin{align}
    \int_{-\pi}^{\pi} d \phi e^\sigma \left[ f_L^{(n)*}(\phi) f_L^{(m)}(\phi)(1+ \Delta_{\tau_\pi, \tau_0})\right]=\delta_{mn}, \nonumber\\
     \label{NonSSBNorm}\\
    \int_{-\pi}^{\pi} d \phi e^\sigma \left[ f_R^{(n)*}(\phi) f_R^{(m)}(\phi)\right]=\delta_{mn}, \nonumber
\end{align}
where we have defined the operator $\Delta_{\tau_\pi, \tau_0} \equiv \frac{2}{kr_c}(\tau_\pi \delta(|\phi|-\pi)+\tau_0 \delta(\phi))$, and $\delta_{mn}$ is just the usual Kronecker delta symbol. In order to obtain the mass terms, we must have
\begin{align}
    \int_{-\pi}^{\pi} d \phi \left[ f_L^{(m)*}(\phi)(\partial_{\phi}f_R^{(n)}(\phi) + r_c sgn(\phi) \nu k f_R^{(n)}(\phi)) \right] &= r_c m_n \delta_{m n}, \nonumber\\
     \\
    \int_{-\pi}^{\pi} d \phi \left[ f_R^{(m)*}(\phi)(\partial_{\phi}f_L^{(n)}(\phi) - r_c sgn(\phi) \nu k f_L^{(n)}(\phi)) \right] &= -r_c m_n \delta_{m n}.\nonumber
\end{align}

The kinetic and mass terms of Eq.(\ref{GoalActionNonSSB}) imply the following equations of motion
\begin{align}
    (\partial_{\phi}+r_c \textrm{sgn}(\phi)\nu k )f_R^{(n)} &= r_c m_n (1+ \Delta_{\tau_\pi,\tau_0})f_L^{(n)}, \nonumber\\
     \label{NonSSBR}\\
    (\partial_{\phi} - r_c \textrm{sgn}(\phi)\nu k )f_L^{(n)} &= r_c m_n f_R^{(n)}. \nonumber
\end{align}
These equations of motion then yield the following solutions for $f^{(n)}_{L,R}$
\begin{align}
    f_L^{(n)}(\phi) &= \frac{e^{\sigma/2}}{N_n}\zeta_{\frac{1}{2}-\nu}(z_n), \nonumber\\
    \label{fnL}\\
    f_R^{(n)}(\phi) &= \frac{-sgn(\phi) e^{\sigma/2}}{N_n}\zeta_{-\frac{1}{2}-\nu}(z_n). \nonumber
\end{align}
Here we define the function $\zeta_{q}(z_n)$ as
\begin{equation}\label{zetaDef}
    \zeta_q(z_n) \equiv \alpha_n J_q(z_n)+\beta_n Y_q (z_n)
\end{equation}
where $J_q(x)$ and $Y_q(x)$ are the order-$q$ Bessel functions of the first and second kind, respectively. The variable $z_n \equiv \frac{m_n}{k}e^\sigma$ is a function of $\phi$ and the mass of the KK mode described by the index $n$, given by $m_n$. The normalization constant $N_n$ is given by
\begin{equation}\label{Normalization}
\int_{-\pi}^{\pi} d \phi (1+\Delta_{\tau_\pi,\tau_0})e^\sigma f_L^{(n)}(\phi)f_L^{(m)}(\phi)=\int_{-\pi}^{\pi} d \phi e^\sigma f_R^{(n)}(\phi)f_R^{(m)}(\phi)=\delta^{n m}.
\end{equation}
Finally, the constants $\alpha_n$ and $\beta_n$ in Eq.(\ref{zetaDef}) are given by boundary conditions on the UV-brane (determined by integrating Eqs. (\ref{NonSSBR}) over an infinitesimal interval of $\phi$ about $\phi=0$)
\begin{eqnarray}
    \alpha_n \equiv Y_{-\frac{1}{2}-\nu}(\epsilon x_n)+\tau_0 \epsilon x_n Y_{\frac{1}{2}-\nu}(\epsilon x_n), \label{alphan}\\
    \beta_n \equiv -[J_{-\frac{1}{2}-\nu}(\epsilon x_n)+\tau_0 \epsilon x_n J_{\frac{1}{2}-\nu}(\epsilon x_n)].\nonumber
\end{eqnarray}
Here, it is convenient to employ the value of $z_n$ evaluated at the TeV-brane ($\phi=\pi$), \ie, $x_n$. Then, $z_n$ evaluated at the UV-brane is given by $\epsilon x_n$, where $\epsilon \equiv e^{-kr_c \pi}$. To find the set of allowed values of $x_n$, and hence the masses of KK tower modes, we must find the roots of the TeV-brane boundary condition equation
\begin{equation}\label{basicBoundary}
\zeta_{-\frac{1}{2}-\nu}(x_n)-\tau_\pi x_n \zeta_{\frac{1}{2}-\nu}(x_n) = 0,
\end{equation}
with $x_n$ being the roots of this equation. The masses of the particles in the KK tower are then $m_n=x_n k \epsilon$. It should also be noted that, in the absence of spontaneous symmetry breaking, the even field also possesses a massless zero-mode solution, given by the (normalized) wave equation,
\begin{equation}
f_L^{(0)}(\phi)=\left( \sqrt{\frac{kr_c}{2}\frac{1+2\nu}{(1+(1+2\nu)\tau_\pi)e^{(1+2\nu)kr_c \pi}-(1-(1+2\nu)\tau_0)}}\right) e^{\nu \sigma}.
\label{fnL0}
\end{equation}

In this work, we will determine which values of $\tau_\pi$, $\tau_0$, and $\nu$ are permitted based on a set of physical conditions. First, all solutions of Eq.(\ref{basicBoundary}) (\ie, the roots $x_n$), which yields the mass spectrum of the Kaluza-Klein tower, must be real. Otherwise, the theory would predict the existence of phenomenologically unviable states (tachyons, for purely imaginary solutions, or fermions with complex masses squared, for general complex solutions). Secondly, we require the absence of so-called ``ghost" states, which are states with negative probability, as indicated by negative values for the square of the Kaluza-Klein mode's normalization. As is standard practice in the literature \cite{dillon,rizzoHD,Aguila:2003,Davoudiasl:2002ua,Davoudiasl:2003zt}, we limit our discussion of the existence of ghost states to the zero-mode $f_L^{(0)}$, which yields the condition
\begin{equation}\label{NoGhostsBasic}
\frac{1+2 \nu}{(1+(1+2\nu)\tau_\pi)\epsilon^{-(\frac{1}{2}+\nu)}-(1-(1+2\nu)\tau_0)\epsilon^{(\frac{1}{2}+\nu)}}>0,
\end{equation}
to avoid ghosts.

\section{Analysis}

Having set up the basic machinery, and in particular established the conditions in Eqs. (\ref{basicBoundary}) and (\ref{NoGhostsBasic}) to judge the physicality of a point in parameter space, we begin our analysis by addressing the specific case that frequently bedevils bulk fields in theories of extra dimensions, namely the existence of tachyonic (purely imaginary) Kaluza-Klein masses \cite{Aguila:2003}. In the following sections, we address the conditions under which tachyonic modes do not appear while the no-ghost condition of Eq.(\ref{NoGhostsBasic}) is simultaneously satisfied. Later, we demonstrate that in the absence of spontaneous symmetry breaking, Kaluza-Klein modes can only appear with purely real or purely imaginary masses, indicating that our analysis here, where ghosts and purely imaginary masses are avoided, produces a complete picture of the allowable parameter space of the model.

\subsection{Study of the Boundary Value Equation}\label{Setup}

First, in an effort to simplify the algebra, we introduce a slightly more convenient fermion localization parameter, $\eta$, by defining
\begin{equation}\label{etaDef}
\eta \equiv -(\frac{1}{2}+\nu).
\end{equation}
The no-ghost condition Eq.(\ref{NoGhostsBasic}) then becomes
\begin{equation}\label{Ghosts}
|N_0|^2 \equiv \frac{-2 \eta}{(1-2 \eta \tau_\pi)\epsilon^\eta-(1+2 \eta \tau_0)\epsilon^{-\eta}}>0.
\end{equation}
Meanwhile, the boundary value equation in Eq.(\ref{basicBoundary}) evaluated on the imaginary line becomes (where we have taken $x \rightarrow i x$ in Eq.(\ref{basicBoundary}), implying that $x$ in the expression below is real) 
\begin{align}
\mathcal{J}(\eta, \tau_\pi, i x) \mathcal{Y}(\eta, -\tau_0, i \epsilon x)-\mathcal{Y}(\eta, \tau_\pi, i x)\mathcal{J}(\eta, -\tau_0, i \epsilon x)=0 \nonumber,\\
\mathcal{J}(\eta, \tau, i x) \equiv J_{\eta}(i x) - i x \tau J_{1+\eta}(i x) \label{BesselBoundaryEq},\\
\mathcal{Y}(\eta, \tau, i x) \equiv Y_{\eta}(i x) - i x \tau Y_{1+\eta}(i x) \nonumber.
\end{align}
If this equation has a root at some $i x$, then, it denotes the existence of a KK mode with a tachyonic mass proportional to this value of $i x$. The expression can be expanded in a double power series using the identities,
\begin{align}
J_\eta(i x)=\bigg( \frac{i x}{2} \bigg)^\eta \sum_{k=0}^{\infty}\bigg( \frac{x}{2} \bigg)^{2k}\frac{1}{k! \Gamma(1+k+\eta)} \label{BesselIDs},\\
Y_\eta(i x)=\cot(\eta \pi)J_\eta(i x)-\csc(\eta \pi)J_{-\eta}(i x) \nonumber,
\end{align}
and then takes the general form (which we define as $f(x)$)
\begin{align}
f(x) &\equiv \frac{1}{\pi \eta} \sum_{k=0}^{\infty}\sum_{j=0}^{k}\bigg( \frac{x}{2}\bigg)^{2 k}\frac{\epsilon^{2 j}}{k!}\binom{k}{j} \bigg[ \frac{\epsilon^\eta(1+2(k-j-\eta)\tau_\pi)(1-2 j \tau_0)\Gamma(1-\eta)\Gamma(1+\eta)}{\Gamma(1+k-j-\eta)\Gamma(1+j+\eta)} \nonumber\\
&- \frac{\epsilon^{-\eta}(1-2 (j-\eta) \tau_0)(1+2(k-j) \tau_\pi)\Gamma(1-\eta)\Gamma(1+\eta)}{\Gamma(1+k-j+\eta)\Gamma(1+j-\eta)}\bigg]=0. \label{GeneralBoundaryEq}
\end{align}

We now assume that $x$ is not hierarchically large (\ie, $x \ll \epsilon^{-1}$). This is motivated by the fact that the RS model is assumed to be a low-energy approximation of some UV-complete theory, and hence subject to an ultraviolet cutoff. Otherwise, a hierarchically large tachyonic root would appear, corresponding to a KK mode with a tachyonic mass near the 4-dimensional Planck scale, jeopardizing the model's validity. In fact, for practical purposes the ultraviolet cutoff must be substantially below the 4-dimensional Planck scale; as noted in \cite{Randall:1999ee}, the UV cutoff for these theories should be reasonably close to the scale $k \epsilon$, to avoid fine tuning in loop corrections to the weak scale that the Randall-Sundrum model is specifically constructed to prevent. Taking $x \ll \epsilon^{-1}$, we see that many terms that are suppressed by powers of $\epsilon^2 x^2$ or higher in $f(x)$ in Eq.(\ref{GeneralBoundaryEq}) can be dropped (which corresponds to neglecting all but the $j=0$ term of the expansion), leading to the following power series expression for the tachyonic root equation
\begin{align}
f(x) \approx \frac{1}{\pi \eta} \sum_{k=0}^{\infty}\bigg( \frac{x}{2}\bigg)^{2 k}\frac{1}{k!}\bigg[ \frac{\epsilon^\eta(1+2(k-\eta)\tau_\pi)\Gamma(1-\eta)}{\Gamma(1+k-\eta)} \nonumber\\
-\frac{\epsilon^{-\eta}(1+2 \eta \tau_0)(1+2 k \tau_\pi)\Gamma(1+\eta)}{\Gamma(1+k+\eta)}\bigg]=0. \label{BoundaryEq}
\end{align}

For all practical purposes, except for the special case when $\tau_0=-1/(2 \eta)$ and $\eta \gsim 0.1$ (which shall be treated separately below), this expansion is sufficient to establish the existence or absence of non-hierarchically-large tachyonic roots for the fermionic KK modes. Interestingly, we note that the $x^0$ term in $f(x)$ is equal to $-2/|N_0|^2$, with $|N_0|^2$ given by Eq.(\ref{Ghosts}). Since $|N_0|^2$ must be positive to avoid ghosts, we see that for any physically valid point in parameter space, the $x^0$ term in Eq.(\ref{BoundaryEq}) is correspondingly negative. So, to avoid ghosts, we see that $f(0)<0$. Now, we consider the possibility that $f(x)>0$ at some $x>0$ (because $f(x)$ is even in $x$, this may be assumed without loss of generality). If there are no ghost states, we then know that $f(0)<0$. So, by the intermediate value theorem, there must exist a point $0<y<x$ such that $f(y)=0$, satisfying Eq.(\ref{BoundaryEq}) and indicating the existence of a tachyonic KK mode. If for some set of values of $\eta$, $\tau_\pi$, and $\tau_0$ there exists a real $x$ such that $f(x)>0$, then this particular set of $\eta$, $\tau_\pi$, and $\tau_0$ values are unphysical: If ghost states are avoided by satisfying Eq.(\ref{Ghosts}), then there must exist a tachyonic root given by the solution to Eq.(\ref{BoundaryEq}), while if Eq.(\ref{Ghosts}) is not satisfied, the point is physically disallowed due to the existence of ghost states. Thus, in order to $\textit{avoid}$ both tachyonic roots and ghost states, one must always have
\begin{align}
f(x) = \frac{1}{\pi \eta} \sum_{k=0}^{\infty}\bigg( \frac{x}{2}\bigg)^{2 k}\frac{1}{k!}\bigg[ \frac{\epsilon^\eta(1+2(k-\eta)\tau_\pi)\Gamma(1-\eta)}{\Gamma(1+k-\eta)} \nonumber\\
-\frac{\epsilon^{-\eta}(1+2 \eta \tau_0)(1+2 k \tau_\pi)\Gamma(1+\eta)}{\Gamma(1+k+\eta)}\bigg]<0 \label{NoTachyonsGeneral}
\end{align}
for all real and non-hierarchically large $x$. For the sake of definiteness, we define ``non-hierarchically large" as being below some cut-off, which we denote as $x_{max}$. As we shall later see, the boundaries of the allowed parameter space are only weakly dependent on $x_{max}$, so that a specific choice for the value of $x_{max}$ is not overly consequential for our final results. In our analysis below, we will examine this equation region by region, covering the RS parameter space.

In our analysis, it shall at times be useful to have an approximate form of the boundary value equation for large, but not hierarchically large, $x$ (\ie, $1 \ll x \ll \epsilon^{-1}$). To find this expression, we employ the asymptotic form of the modified Bessel function of the first kind, $I_\eta(x) \equiv i^\eta J_\eta(x)$
\begin{equation}
I_\eta (x) \approx \frac{e^x}{\sqrt{2 \pi x}} \sum_{k=0}^{\infty}\frac{\Gamma(\frac{1}{2}+k+\eta)\Gamma(\frac{1}{2}+k-\eta)}{k! \Gamma(\frac{1}{2}+\eta)\Gamma(\frac{1}{2}-\eta)}\bigg( \frac{1}{2 x} \bigg)^{k}.
\end{equation}
Here, terms proportional to $e^{-x}$ have been dropped, rendering this expression only valid for large $x$. This expansion yields the following recasting of Eq.(\ref{NoTachyonsGeneral})
\begin{multline}\label{LargeXExpansion}
f(x) \approx \frac{e^x}{\eta \pi^{3/2} \sqrt{2 x}}\bigg[ \bigg( \frac{\epsilon x}{2} \bigg)^{\eta}\Gamma(1-\eta)-(1+2\eta \tau_0)\bigg( \frac{\epsilon x}{2} \bigg)^{-\eta} \Gamma(1+\eta)\bigg]\\
\\
\times \sum_{k=0}^{\infty}\bigg( \frac{1}{2 x} \bigg)^{k}\bigg[ \frac{\Gamma(\frac{1}{2}+k-\eta)\Gamma(\frac{1}{2}+k+\eta)}{k! \Gamma(\frac{1}{2}-\eta)\Gamma(\frac{1}{2}+\eta)}+\frac{x \tau_\pi \Gamma(-\frac{1}{2}+k-\eta)\Gamma(\frac{3}{2}+k+\eta)}{k!\Gamma(-\frac{1}{2}-\eta)\Gamma(\frac{3}{2}+\eta)}\bigg]<0.
\end{multline}

We will now examine each section of the parameter space, one-by-one.

\subsection{Fermions Near the TeV-brane ($\eta\lesssim -0.1$)}\label{TeVFermions}

Having rewritten our boundary value equation, we now address the case where the fermion is localized ``close" to the TeV-brane ($\eta$ is large and negative), far enough from $\eta=0$ so that the $\epsilon^{-\eta}$ terms can be safely ignored relative to the $\epsilon^{\eta}$ terms in Eq.(\ref{BoundaryEq}). In practice, a numerical investigation indicates that the condition for this approximation to be valid is roughly $\eta\lesssim -0.1$. In this case, $\epsilon^{-\eta}\lesssim 0.02$, so that a $\sim 4 \times 10^{-4}$ level suppression of the $\epsilon^{-\eta}$ terms occurs relative to the $\epsilon^{\eta}$ terms. Assuming natural (magnitude $<50$) values for $\tau_0$ and $\tau_\pi$, this leads to at most a $\sim 1\%$ discrepancy between the value of $f(x)$ with the $\epsilon^{-\eta}$ terms dropped versus being included. The condition to avoid tachyons then becomes (noting that $\eta<0$ here)
\begin{equation}\label{TeVTachyon}
\tilde{f}(x)|_{TeV} \equiv (\eta \pi) f(x)|_{TeV} \approx \sum_{k=0}^{\infty}\bigg( \frac{x}{2}\bigg)^{2 k}\frac{1}{k!}\bigg[ \frac{\epsilon^\eta(1+2(k-\eta)\tau_\pi)\Gamma(1-\eta)}{\Gamma(1+k-\eta)}\bigg]>0.
\end{equation}
Here, we have defined $\tilde{f}(x)$ as $f(x)$ multiplied by the (negative) value $\eta \pi$, in order to avoid sign confusion later on. Meanwhile, the condition to avoid ghost states simplifies to
\begin{equation}\label{TeVGhost}
1-2\eta \tau_\pi >0.
\end{equation}

First, we consider the case $\tau_\pi \geq 0$. Recalling that $\eta<0$, we see that the no-ghost condition Eq.(\ref{TeVGhost}) is automatically satisfied. We now note that, when $\eta<0$, both $1-\eta$ and $1+k-\eta$ (for some natural number $k$) are positive, and as a result, the quantity $\Gamma(1-\eta)/\Gamma(1+k-\eta)$ is also positive. Meanwhile, since $1-2\eta \tau_\pi > 0$ and $k \tau_\pi$ is also positive, we observe that the coefficient of each $(x/2)^{2k}$ term in $\tilde{f}(x)$ is also positive. Thus we conclude that Eq.(\ref{TeVTachyon}) is always satisfied in this regime when $\tau_\pi \geq 0$, indicating that this region of parameter space avoids both tachyons and ghosts, and is hence physically allowed.

Now, we consider the opposite case where $\tau_\pi<0$. As was found in the case where $\tau_\pi \geq 0$, the ratio $\Gamma(1-\eta)/\Gamma(1+k-\eta)$ remains positive. However, as $k$ gets large, the $k\tau_\pi$ term in the coefficients of Eq.(\ref{TeVTachyon}) will dominate the numerator, and since $\tau_\pi<0$, this results in the existence of an infinite series of negative terms in Eq.(\ref{TeVTachyon}) (\ie, all terms after some minimum index $k$). Because the infinite series of negative terms is proportional to large powers of $x$, Eq.(\ref{TeVTachyon}) must eventually become negative at large $x$, since these higher-order terms will dominate the expansion in that regime. This leads to a violation of the condition to simultaneously avoid tachyons and/or ghosts, physically disallowing this region of parameter space.

In the region where $\tau_\pi<0$, there also exists a single special case that requires individul attention, namely, when $\tau_\pi=1/(2 \eta)$. In this scenario, instead of Eq.(\ref{TeVTachyon}), the general condition Eq.(\ref{NoTachyonsGeneral}) becomes
\begin{equation}\label{TeVTachyonSpecial}
    -(1+2 \eta \tau_0)\epsilon^{-\eta}+\sum_{k=1}^{\infty} \left( \frac{x}{2} \right)^{2 k} \frac{1}{k!} \frac{\epsilon^{\eta}k \Gamma(1-\eta)}{\eta \Gamma(1+k-\eta)} > 0,
\end{equation}
where we have substituted the value $\tau_\pi = 1/(2 \eta)$ into our expression for $\tilde{f}(x)$, and noted that, because the $x^0$ term in the above expansion has no part proportional to $\epsilon^{\eta}$, we cannot omit the part proportional to $\epsilon^{-\eta}$. Meanwhile, because the $\epsilon^{\eta}$ contribution in Eq.(\ref{Ghosts}) (the condition to avoid ghosts), is equal to 0, we obtain a different no-ghost condition from that of Eq.(\ref{TeVGhost}), namely
\begin{equation}
    1+2 \eta \tau_0< 0.
\end{equation}
Even under these new conditions, however, we see that all higher-order ($x^2$ or higher) terms in Eq.(\ref{TeVTachyonSpecial}) are negative, because $\eta < 0$ and, as before, $\Gamma(1-\eta)$ and $\Gamma(1+k-\eta)$ are positive. So, even if the $x^0$ term of Eq.(\ref{TeVTachyonSpecial}) is positive, satisfying the no-ghost condition, all subsequent terms in this expansion must be negative, eventually forcing Eq.(\ref{TeVTachyonSpecial}) to be violated at some $x$. In this special case, as for the general region $\tau_\pi<0$, then, tachyonic roots and ghost states cannot be simultaneously avoided.

In summary, we find that when $\eta\lesssim-0.1$, which indicates that a fermion is localized close to the TeV-brane, the general condition required to prevent the existence of tachyons and ghost states is $\tau_\pi \geq 0$.

\subsection{Fermions Near the UV-brane ($\eta \protect\gsim 0.1$)}\label{UVFermions}

Having dealt with the case where fermions are localized close to the TeV-brane, we now address the opposite extreme, in which fermions reside close to the UV-brane, now given by the corresponding condition $\eta \gsim 0.1$. Notably, results derived here are also applicable to bulk graviton fields, where their tachyonic spectra are given by Eq.(\ref{BesselBoundaryEq}) when $\eta=1$, and the fermion brane terms are replaced by their graviton counterparts \footnote{For a detailed discussion of brane-localized terms for graviton fields in the RS model, we refer the reader to \cite{Davoudiasl:2003zt}.}. There are two scenarios to consider here, one in which $(1+ 2 \eta \tau_0) = 0$ and one in which $(1+2 \eta \tau_0) \neq 0$. We shall address the latter case first, since it is simpler, and then move on to the specialized region where $(1+2 \eta \tau_0)=0$.

\subsubsection{The Case $(1+2 \eta \tau_0) \neq 0$}

Assuming $(1+2 \eta \tau_0) \neq 0$, Eq.(\ref{NoTachyonsGeneral}) reduces in the UV-brane localized limit to
\begin{equation}\label{UVTachyon}
\sum_{k=0}^{\infty}\bigg( \frac{x}{2}\bigg)^{2 k}\frac{1}{k!}\bigg[\frac{\epsilon^{-\eta}(1+2 \eta \tau_0)(1+2 k \tau_\pi)\Gamma(1+\eta)}{\Gamma(1+k+\eta)}\bigg]>0,
\end{equation}
while the no-ghost condition Eq.(\ref{Ghosts}) becomes
\begin{equation}\label{UVGhost}
1+2 \eta \tau_0>0.
\end{equation}
Note that since $\eta>0$, the quantity $\Gamma(1+\eta)/\Gamma(1+k+\eta)>0$ for any natural number $k$. Furthermore, Eq.(\ref{UVGhost}) then requires that $(1+2\eta \tau_0)\Gamma(1+\eta)/\Gamma(1+k+\eta)$ be positive. Thus, the sign of the $k^{th}$ term in the power series of Eq.(\ref{UVTachyon}) is determined by the sign of $(1+2k \tau_\pi)$. If $\tau_\pi \geq 0$, this will then result in every term of the power series having a positive coefficient, automatically satisfying the tachyon-free condition of Eq.(\ref{UVTachyon}). However, if $\tau_\pi <0$, then for some sufficiently large $k$, $1+2 k \tau_\pi$ becomes negative and remains negative for all subsequent terms in the expansion. As a result, the tachyon-free condition Eq.(\ref{UVTachyon}) will eventually be violated, indicating the existence of a tachyonic root. So, in the case where $1+2 \eta \tau_0 \neq 0$, the conditions required to avoid tachyons and ghost states are simply $\tau_\pi \geq 0$ and $1 + 2 \eta \tau_0 >0$.

Notably, while our treatment here is based on the Kaluza-Klein decomposition of a bulk fermion field, the resultant expressions for the bulk profile of the massless zero-mode and the boundary value equation for Kaluza-Klein states apply equally well to bulk gravition fields, as long as the localization parameter $\eta$ is set to 1, and the fermion brane-localized kinetic terms $\tau_\pi$ and $\tau_0$ are substituted for corresponding brane-localized curvature terms $\delta_\pi$ and $\delta_0$ (these are defined analogously to the fermion brane-localized kinetic terms, with the only exception being that they are coefficients of 4-dimensional scalar curvature terms, rather than 4-dimensional fermion kinetic terms) \cite{Davoudiasl:2003zt}. The restrictions on the parameter space for gravitons are then trivially derived by setting $\eta=1$ and substituting $\delta_{0,\pi}$ for $\tau_{0,\pi}$ in Eqs. (\ref{UVTachyon}) and (\ref{UVGhost}). However, as noted in \cite{dillon,Davoudiasl:2003zt}, the existence of the radion field for bulk gravitons requires that, to avoid radion ghost states, the parameter $\delta_\pi$ must also follow the bound $\delta_\pi \leq 1$. While it has been noted that Higgs-radion mixing may relax this bound somewhat \cite{dillon}, a full exploration of this bound goes beyond the scope of this analysis, so we restrict our discussion to quoting the $0 \leq \delta_\pi \leq 1$ result.

\subsubsection{The Case $(1+2 \eta \tau_0) = 0$}\label{UVSpecial}
The case where $1+ 2 \eta \tau_0 = 0$ and $\eta \gsim 0.1$ is a small, but non-trivial, region of parameter space, where the analysis is complex enough to merit separate treatment. It should be noted that this ``line" in the $\tau_0$-$\eta$ plane is technically an approximation of an extremely narrow band in this plane, corresponding to where the term proportional to $(1+2 \eta \tau_0)$ in Eq.(\ref{GeneralBoundaryEq}), which is normally dominant for UV-brane localized fermions, becomes small enough to be subordinate to other terms. However, in the $\eta$ region we consider here, even where this band is thickest (at $\eta\approx 0.1$, where the subdominant $\epsilon^{\eta}$ term in Eq.(\ref{GeneralBoundaryEq}) is least suppressed compared to the $\epsilon^{\eta}$ term), the $\epsilon^{-\eta}$ term in Eq.(\ref{GeneralBoundaryEq}) only becomes subordinate to other terms in the expansion if $|1+ 2 \eta \tau_0| \lesssim O(10^{-3})$. Given how narrow the region of $1+2 \eta \tau_0$ values must be in order to invalidate our analysis in the previous section, we restrict our discussion here to the line $1+2 \eta \tau_0=0$. Notably, the contributors to the power series in Eq.(\ref{GeneralBoundaryEq}) proportional to $\epsilon^{-\eta}$ are now suppressed by at least $O(\epsilon^2)$. Taking the leading-order non-trivial terms for both the $\epsilon^{\eta}$ and $\epsilon^{-\eta}$ contributions in Eq.(\ref{GeneralBoundaryEq}) leads to the following condition to avoid tachyonic states (where $\tau_0=-\frac{1}{2 \eta}$ has been employed)
\begin{align}
g(x) &\equiv \sum_{k=0}^{\infty} \bigg( \frac{x}{2} \bigg)^{2 k} \frac{\epsilon^\eta}{k!} \bigg[\frac{(1+2(k - \eta)\tau_\pi)\Gamma(1-\eta)}{\Gamma(1+k-\eta)} \nonumber\\
&-\frac{\epsilon^{2(1-\eta)} k (1+2(k-1)\tau_\pi)\Gamma(1+\eta)}{\eta(1-\eta)\Gamma(k+\eta)} \bigg]<0. \label{UVTachyonSpecial}
\end{align}
In the same limit, to avoid ghosts, we must also require
\begin{equation}\label{UVGhostSpecial}
1-2\eta \tau_\pi <0 \rightarrow \tau_\pi>\frac{1}{2 \eta}>0.
\end{equation}

We first consider the region where $\eta$ is far enough below unity that the $\epsilon^{2 (1-\eta)}$ term in Eq.(\ref{UVTachyonSpecial}) may be safely ignored, and in keeping with our procedures elsewhere in this analysis, this region is taken to be approximately $\eta \lesssim 0.9$. Then, Eq.(\ref{UVTachyonSpecial}) reduces to the form
\begin{equation}
g(x) \approx \sum_{k=0}^{\infty}\bigg( \frac{x}{2} \bigg)^{2 k} \frac{\epsilon^\eta}{k!}\bigg[ \frac{(1+2(k-\eta)\tau_\pi)\Gamma(1-\eta)}{\Gamma(1+k-\eta)} \bigg]<0.
\end{equation}
Here, to avoid ghost states, $\tau_\pi>\frac{1}{2 \eta}>0$ as above, so that $1-2 \eta \tau_\pi$ is negative. As $k$ grows large, the $k \tau_\pi$ term in the expression $1+2(k-\eta)\tau_\pi$ will come to dominate the numerator, and since $\tau_\pi > 0$, this term will have a positive value. In addition, because we are considering the region $\eta<1$, both $\Gamma(1-\eta)>0$ and $\Gamma(1+k-\eta)>0$. Therefore, starting at some initial $k_0$, $g(x)$ will have an infinite number of $x^{2k}$ terms with positive coefficients. At large $x$, these terms will eventually force $g(x)$ to become positive, violating the condition in Eq.(\ref{UVTachyonSpecial}). Hence, when $(1+2 \eta \tau_0)=0$, the region $0.1 \lesssim \eta \lesssim 0.9$ is physically $\textit{disallowed}$.

Next, we consider the region $\eta \gsim 1.1$, at which point the $\epsilon^{2(1-\eta)}$ terms in $g(x)$ dominate the other pieces of the expansion. This reduces Eq.(\ref{UVTachyonSpecial}) to the form
\begin{equation}\label{UVTachyonSpecial4}
g(x) \approx \epsilon^\eta(1-2\eta \tau_\pi)+\sum_{k=1}^{\infty}\bigg( \frac{x}{2} \bigg)^{2 k}\frac{(-1)\epsilon^{2-\eta}}{(k-1)!}\frac{(1+2(k-1)\tau_\pi)\Gamma(1+\eta)}{\eta (1-\eta) \Gamma(k+\eta)}<0.
\end{equation}
As $k$ gets large, the dominant part of the coefficient of the $\left( \frac{x}{2} \right)^{2k}$ term becomes
\begin{equation}
\frac{-\epsilon^{2-\eta}}{(k-1)!}\frac{2k\tau_\pi\Gamma(1+\eta)}{\eta(1-\eta)\Gamma(k+\eta)}.
\end{equation}
Now, recall that in order to avoid ghosts, $\tau_\pi>0$. Since we are working in a region where $\eta>1$, so that the factor $1-\eta<0$, a brief inspection shows that the coefficients for the $(x/2)^{2k}$ terms are all positive in the limit of large $k$. This spawns an infinite number of high-order terms in Eq.(\ref{UVTachyonSpecial4}) which contribute positively to the value of $g(x)$, implying that $g(x)$ will eventually become positive and generate a tachyonic root. Therefore, in the region $\eta \gsim 1.1$, we again cannot simultaneously avoid tachyonic roots and ghosts.

Finally we consider the remaining region $0.9 \lesssim \eta \lesssim 1.1$, where we see that we can no longer neglect terms suppressed by either $\epsilon^{\eta}$ or $\epsilon^{2-\eta}$. Keeping these terms, the condition to avoid tachyons is given by Eq.(\ref{UVTachyonSpecial}). In this case, it is well within the realm of possibility that the limit of $g(x)$ as $x \rightarrow \infty$ is negative, meaning that unlike the other $\eta$ values we have examined above, this region cannot be easily dismissed as yielding tachyonic roots. In particular, we can consider the subregion of this piece of parameter space where $1-\eta < 0$. It can be shown that eventually the $k \tau_\pi\Gamma(1-\eta)/\Gamma(1+k-\eta)$ term dominates this expansion for sufficiently large $k$. In this case, because $-0.1 \lsim 1-\eta<0$ (which in turn implies that $\Gamma(1-\eta)<0$) and $\tau_\pi>0$, naively we observe that the eventual behavior of the expansion should trend towards negative infinity in this regime. To determine if this naive analysis is correct, we probe this small region of parameter space numerically. In practice, we are most interested in the potential existence of positive values of $g(x)$ below a reasonable cutoff (past which we assume the existence of a tachyonic root to be an artifact of the RS model being a low energy effective theory). We take this cutoff to be $x_{max}=500$. 

To more easily numerically examine $g(x)$, we turn to its power series expression. Naively, truncating any power series where $|x|>1$ would appear to be unwise, since higher-order terms in $x$ will generally contribute more to the value of the expansion than their lower-order counterparts. However, this is predicated on the assumption that the coefficients of higher-order terms in $x$ are of comparable magnitude to those of lower-order terms, which is not the case here for $g(x)$. To see this clearly, we define the quantities $A_k$ and $B_k$ such that
\begin{align}
A_k &\equiv \frac{\epsilon^{\eta} (1+2(k-\eta)\tau_\pi)\Gamma(1-\eta)}{k!\Gamma(1+k-\eta)},\\
B_k &\equiv \frac{\epsilon^{2-\eta}(1+2(k-1)\tau_\pi)\Gamma(1+\eta)}{\eta(1-\eta) k! \Gamma(k+\eta)}. \nonumber
\end{align}
Next we define the functions $a(x)$ and $b(x)$ as
\begin{align}
    a(x) &\equiv \sum_{k=0}^{\infty}A_k \bigg( \frac{x}{2} \bigg)^{2 k},\\
    b(x) &\equiv \sum_{k=0}^{\infty}B_k \bigg( \frac{x}{2} \bigg)^{2 k}, \nonumber
\end{align}
so that one may then rewrite Eq.(\ref{UVTachyonSpecial}) as $g(x) = a(x)+b(x)$. Now, to validate the accuracy of truncating the series expansion of $g(x)$, we must determine whether or not for some $x$ there exists a value $k_0$ such that, for any $k \geq k_0$, the term $A_k \left( \frac{x}{2} \right)^{2k}$ is larger in magnitude than the term $A_{k+1} (x/2)^{2(k+1)}$ in $a(x)$, and correspondingly for the expansion terms in $b(x)$. If this is the case, then it is reasonable to truncate the series for $g(x)$ comfortably past $k_0$, so that the terms of the series omitted by truncation are all numerically insignificant. We are specifically concerned with terms where $k$ is large (namely, where the terms proportional to $k \tau_\pi$ in $A_k$ and $B_k$ dominate the values of these terms), if only because it is a simple enough matter to include the finite number of terms in the power expansion of $g(x)$ where $k$ is not large. In the limit where $k$ is large, $A_k$ and $B_k$ become
\begin{align}
    A_k &\approx \frac{\epsilon^\eta(2 k \tau_\pi) \Gamma(1-\eta)}{k! \Gamma(1+k-\eta)},\\
    B_k &\approx \frac{\epsilon^{2-\eta}(2 k \tau_\pi) \Gamma(1+\eta)}{\eta(1-\eta) k! \Gamma(k+\eta)}. \nonumber
\end{align}
Taking ratios of successive terms of $a(x)$ and $b(x)$ then yields
\begin{align}
    \frac{\left( \frac{x}{2} \right)^{2 (k+1)}A_{k+1}}{\left( \frac{x}{2} \right)^{2 k}A_{k}} &= \bigg( \frac{x}{2} \bigg)^{2} \frac{1}{k (1+k-\eta)} \approx \bigg( \frac{x}{2} \bigg)^{2} \frac{1}{k^2},\\
    \frac{\left( \frac{x}{2} \right)^{2 (k+1)}B_{k+1}}{\left( \frac{x}{2} \right)^{2 k}B_{k}} &= \bigg( \frac{x}{2} \bigg)^{2} \frac{1}{k (k+\eta)} \approx \bigg( \frac{x}{2} \bigg)^{2} \frac{1}{k^2}. \nonumber
\end{align}
Thus, we see that for $\frac{x^2}{4} \lsim k^2$ (\ie, $x \lsim 2 k$), the ratio of the $\left( \frac{x}{2} \right)^{2(k+1)}$ term to the $\left( \frac{x}{2} \right)^{2k}$ term in either $a(x)$ or $b(x)$ will be less than unity. This indicates that past a certain $k$ value, higher-order terms in these functions will contribute less to the expansion than lower-order terms. We thus conclude that as long as we select a cutoff value for $k$ large enough so that we can anticipate any higher-order terms in $a(x)$ and $b(x)$ will contribute negligibly at our cutoff $x_{max}$, then $g(x) = a(x)+b(x)$ can be well approximated even when the sum in Eq.(\ref{UVTachyonSpecial}) is truncated.

For our numerical analysis (using Mathematica \cite{mathematica}), we truncate the series at $k_{max} = 500$ (not to be confused with $x_{max}$), neglecting the terms proportional to $x^{2(501)}$ and higher. Based on the suppression of terms in $g(x)$, this should be more than sufficient to faithfully approximate the value of $g(x)$ for any $x<x_{max}=500$, given that according to our preceding analysis, all terms with $k>250$ should contribute progressively less to the value of $g(x)$ than each term with lower $k$ within this region of $x$. Our numerical analysis finds positive maxima for Eq.(\ref{UVTachyonSpecial}) when $x<500$ for all points in the parameter space where $\tau_\pi<50$ and $0.9< \eta < 1.1$. Thus, we that find the region where $\eta \approx 1$ is also disallowed. Hence, the entire region where $\tau_0=-1/(2\eta)$ is disallowed for fermions localized near the UV-brane.

In summary then, the only allowed region of parameter space for UV-brane localized fermions is $\tau_\pi \geq 0$ and $\tau_0 > -1/(2\eta)$. This also implies that for gravitons, the allowed parameter space for the brane-localized curvature terms $\delta_\pi$ and $\delta_0$ (localized on the IR- and UV-brane, respectively) based on our physicality conditions is $0 \leq \delta_\pi \leq 1$ and $\delta_0 > -\frac{1}{2}$, where the bound $\delta_\pi \leq 1$ is required to avoid ghost states for the radion field, which does not exist for UV-localized fermions.

\subsection{The Region $-0.1 \lesssim \eta \lesssim 0.1$}\label{GaugeFermions}

Since $\epsilon \sim O(10^{-16})$ is such a small parameter, a treatment of the potential existence of tachyons and ghost-like states in the simple $\eta \rightarrow 0$ limit is insufficient to explore the full range of results where $\eta$ is too small to ignore either the $\epsilon^{\eta}$ or $\epsilon^{-\eta}$ terms in Eq.(\ref{NoTachyonsGeneral}). As a result, we must instead consider the somewhat larger region where $-0.1 \lesssim \eta \lesssim 0.1$, where our prior practice of neglecting either the $\epsilon^{\eta}$ or $\epsilon^{-\eta}$ terms is no longer valid. In the limit where $\eta$ is non-vanishing, but close to zero, Eq.(\ref{NoTachyonsGeneral}) becomes
\begin{multline}\label{UVCloseTachyon}
f(x) \equiv \sum_{k=0}^{\infty}\bigg( \frac{x}{2} \bigg)^{2 k} \frac{C_k}{k!} \equiv \frac{1}{\pi \eta}\sum_{k=0}^{\infty}\bigg( \frac{x}{2}\bigg)^{2 k}\frac{1}{k!}\bigg[ \frac{\epsilon^\eta(1+2(k-\eta)\tau_\pi)\Gamma(1-\eta)}{\Gamma(1+k-\eta)}\\
-\frac{\epsilon^{-\eta}(1+2 \eta \tau_0)(1+2 k \tau_\pi)\Gamma(1+\eta)}{\Gamma(1+k+\eta)}\bigg]<0,
\end{multline}
while the corresponding condition Eq.(\ref{Ghosts}) to avoid ghosts is now
\begin{equation}\label{UVCloseGhost}
\frac{1}{\eta}(1+2\eta \tau_0)>\frac{1}{\eta}(1-2\eta \tau_\pi)\epsilon^{2 \eta}.
\end{equation}

This region includes an $\eta$ value of particular interest, namely, $\eta=0$. In this case, the fermion bulk wave functions and resulting spectrum become precisely equivalent to those of a bulk gauge field (for a discussion of RS bulk gauge fields, see, for example, \cite{Davoudiasl:2002ua,bulkgauges}), rendering the constraints we derive in this region relevant to analyses involving bulk gauge fields with BLKT's. This holds even in the absence of any such terms for bulk fermions.

As before, we now address the various regions of parameter space under which these equations are satisfied for different values of $\tau_\pi$.

\subsubsection{The Case $\tau_\pi\geq 0$}

We first consider the case where $\tau_\pi \geq 0$. Here we demonstrate that in this regime, the expression $f(x)$ in Eq.(\ref{UVCloseTachyon}) is positive (violating the condition for the simultaneous absence of ghosts and tachyons) for some $x$ below an arbitrary cutoff $x_{max}$ $\textit{if and only if}$ $f(x_{max})>0$. Hence, if $f(x_{max})$ is negative, then $f(x)$ is also negative for $\textit{all } x \textit{ below}$ $x_{max}$. To prove this lemma, first we note that if $f(x_{max})>0$, then continuity of $f(x)$ requires that for some range of $x$ immediately below $x_{max}$, $f(x)$ is positive as well. However, the converse statement, that if $f(x)>0$ for some $x<x_{max}$, then $f(x_{max})$ will be positive, is less trivial. We first note that a necessary condition for $f(x)$ to be positive at some value of $x$ is that at least one coefficient $C_k$ in the expansion of Eq.(\ref{UVCloseTachyon}) be non-negative. Otherwise, all terms would be negative, and it would be impossible to violate the condition in Eq.(\ref{UVCloseTachyon}). We then show that if some $C_{k_0} \geq 0$ for some value of $k_0$, then $C_k>0$ for all $k>k_0$. 

We prove this lemma by contradiction, namely, by demonstrating that for some natural number $k_0$, it is impossible for both the conditions $C_{k_0} \geq 0$ and $C_{k_0+1} \leq 0$ to be satisfied. First, we note that the coefficient $C_{k_0}$ in Eq.(\ref{UVCloseTachyon}) is non-negative provided that
\begin{equation}
\frac{\epsilon^{\eta}(1+2(k_0-\eta)\tau_\pi)\Gamma(1-\eta)}{\eta \Gamma(1+k_0-\eta)} \geq \frac{\epsilon^{-\eta}(1+2 \eta \tau_0)(1+2 k_0 \tau_\pi)\Gamma(1+\eta)}{\eta \Gamma(1+k_0+\eta)}.
\end{equation}
If, however, $C_{k_0+1} \leq 0$, we see that
\begin{equation}
\frac{\epsilon^{\eta}(1+2(k_0+1-\eta)\tau_\pi)\Gamma(1-\eta)}{\eta (1+k_0-\eta)\Gamma(1+k_0-\eta)} \leq \frac{\epsilon^{-\eta}(1+2 \eta \tau_0)(1+2 (k_0+1) \tau_\pi)\Gamma(1+\eta)}{\eta (1+k_0+\eta)\Gamma(1+k_0+\eta)}.
\end{equation}
These two equations may be rewritten (taking advantage of the fact that $\tau_\pi\geq 0$ so that $1+2k_0 \tau_\pi>0$) as
\begin{align}
\frac{\epsilon^{-2\eta}}{\eta}(1+2\eta \tau_0) &\leq \frac{(1+2(k_0-\eta)\tau_\pi)\Gamma(1-\eta)\Gamma(1+k_0+\eta)}{\eta(1+2k_0 \tau_\pi)\Gamma(1+\eta)\Gamma(1+k_0-\eta)} \label{Gauget0Bound1},\\
\nonumber \\
\frac{\epsilon^{-2 \eta}}{\eta}(1+2 \eta \tau_0) & \geq \frac{(1+2(k_0+1-\eta)\tau_\pi)(1+k_0+\eta)\Gamma(1-\eta)\Gamma(1+k_0+\eta)}{\eta(1+2(k_0+1)\tau_\pi)(1+k_0-\eta)\Gamma(1+\eta)\Gamma(1+k_0-\eta)}.\nonumber
\end{align}
These two conditions constrain $\epsilon^{-2 \eta}(1+2 \eta \tau_0)$ to a particular range, and for this range to have finite measure, the right side of the upper expression in Eq.(\ref{Gauget0Bound1}) must be be greater than or equal to the right side of the lower expression. Setting the former expression greater than or equal to the latter, and dividing out the gamma functions from both sides{\footnote{Note that, since $|\eta|<1$ here, these functions may be divided out of any inequality without modifying that inequality's direction.}}, we arrive at the condition
\begin{equation}
\frac{(1+2(k_0+1-\eta)\tau_\pi)(1+k_0+\eta)}{\eta (1+2(k_0+1)\tau_\pi)(1+k_0-\eta)} \leq \frac{(1+2(k_0-\eta)\tau_\pi)}{\eta (1+2 k_0 \tau_\pi)}.
\end{equation}
This can now be further reduced to a quadratic inequality in $\tau_\pi$, given by
\begin{equation}
(1+k_0-\eta)(1+2k_0)\tau_\pi^2+(1+2k_0-\eta)\tau_\pi+\frac{1}{2} \leq 0,
\end{equation}
with a discriminant $\eta^2-1-2k_0$. Notably, when $|\eta|<1$, as is the case in the region we are considering, this discriminant can never be positive, because $k$ is a non-negative integer and therefore $1+2k \geq 1$. Meanwhile, for the same reason, the coefficient of $\tau_\pi^2$, namely $(1+k-\eta)(1+2k)$, is positive. Thus, we see that there is no region in the parameter space we are considering where this quadratic inequality in $\tau_\pi$ can be satisfied. This further implies that there is no region in this space in which there can exist $k_0$ such that $C_{k_0} \geq 0$ and $C_{k_0+1} \leq 0$. By repeatedly applying this lemma, we observe that if $C_{k_0} \geq 0$ for some $k_0$, then $C_k > 0$ for all $k>k_0$. Using this result, we see that if $f(x)>0$ at some $x$, it has a finite number (possibly zero) of lower-order (in $x$) terms that have non-positive coefficients, followed by an infinite number of higher-order terms with positive coefficients. 

Having proved the above lemma, we can now return to our original goal, namely, demonstrating that $f(x)>0$ for some $x< x_{max}$ if and only if $f(x_{max})>0$. After showing that if $f(x_{max})>0$, then there exists an $x<x_{max}$ such that $f(x)>0$, our sole remaining task is to demonstrate the converse. We accomplish this by using our previously derived lemma on the expansion coefficients $C_k$. To begin, we consider the scenario where $f(x)>0$ for some $x<x_{max}$. So, $f(x)$ may be written as
\begin{equation}
f(x) = \sum_{k=0}^{n-1}\bigg( \frac{x}{2} \bigg)^{2k} \frac{C_k}{k!}+\sum_{j=n}^{\infty} \bigg( \frac{x}{2} \bigg)^{2j} \frac{C_j}{j!}>0,
\end{equation}
where here, all $C_k \leq 0$, and all $C_j > 0$, due to our previously proven statement that if any coefficient $C_{k_0}$ is non-negative, then $C_k>0$ holds for all $k>k_0$. So, the expansion of $f(x)$ contains exactly $n$ terms with non-positive coefficients, followed by an infinite number of terms that all have positive coefficients{\footnote{Notably, in the event that $n=0$ (so that all terms in $f(x)$ are positive), the above formula must be modified slightly: The first sum, from $k=0$ to $k=n-1$, will be dropped entirely in this case.}}. Now, observe that
\begin{equation}
\frac{x}{2}f'(x)=\frac{x}{2}\frac{d f(x)}{dx} = \sum_{k=1}^{n-1} \bigg( \frac{x}{2} \bigg)^{2k} \frac{k C_k}{k!}+\sum_{j=n}^{\infty} \bigg( \frac{x}{2} \bigg)^{2j} \frac{j C_j}{j!}.
\end{equation}
We note that because each $C_k \leq 0$, for each $k < n$, and each $C_j > 0$, for each $j \geq n$, it follows that
\begin{equation}\label{DerivativeBound}
\frac{x}{2}\frac{d f(x)}{d x} > n \sum_{k=1}^{n-1} \bigg( \frac{x}{2} \bigg)^{2k} \frac{C_k}{k!}+ n \sum_{j=n}^{\infty} \bigg( \frac{x}{2} \bigg)^{2j} \frac{C_j}{j!}=n (f(x)-C_0).
\end{equation}
We then have two scenarios to consider. If $n > 0$, then because the first $n$ terms in the expansion of $f(x)$ are non-positive, $C_0 \leq 0$, so that we find that $(x/2)f'(x) > n f(x) > 0$, since by construction we have assumed $f(x)>0$. The other scenario, $n=0$, implies that Eq.(\ref{DerivativeBound}) automatically stipulates that $(x/2)f'(x) > 0$. In all cases, $f(x)$ has a positive derivative if $f(x)>0$, indicating that this function is always increasing wherever $f(x)>0$. Therefore, if $f(x)>0$, then $f(x_{max})>f(x)>0$ for any $x< x_{max}$. Hence, in the region where $\tau_\pi \geq 0$, the tachyon-free condition Eq.(\ref{UVCloseTachyon}) is violated for some $x<x_{max}$ if and only if $f(x_{max})>0$.

With this proof in hand, we can now find the region of parameter space that avoids ghosts and tachyons solely by probing the points in parameter space where $f(x_{max}) > 0$, where $x_{max}$ is the cutoff past which we consider tachyonic roots hierarchically large and therefore unphysical artifacts. To probe $f(x)$ at $x_{max}$, we use the asymptotic expansion given in Eq.(\ref{LargeXExpansion}). Keeping only terms proportional to $x$ or $x^0$ (all other terms are suppressed by at least $x^{-1}$) in this expansion, we derive an approximate expression for $f(x_{max})$ given by
\begin{multline}\label{UVCloseAsymptotic}
f(x_{max}) \approx \frac{e^{x_{max}}(1-\tau_\pi(\frac{3}{8}+\eta+\frac{1}{2}\eta^2)+x_{max} \tau_\pi)}{\eta \pi^{3/2} \sqrt{2 x_{max}}}\\
\times \bigg[ \bigg( \frac{\epsilon x_{max}}{2} \bigg)^\eta \Gamma(1-\eta)-\bigg( \frac{\epsilon x_{max}}{2} \bigg)^{-\eta} \Gamma(1+\eta)(1+2 \eta \tau_0)\bigg] < 0.
\end{multline}
Using the fact that the $x_{max} \tau_\pi$ term, which is positive because both $x_{max}$ and $\tau_\pi$ are positive, dominates the sign of the expression in the first line of Eq.(\ref{UVCloseAsymptotic}), we arrive at a condition on $\tau_0$ that assures the entire expression $f(x_{max})$ remains positive, namely,
\begin{equation}\label{t0GaugeLike}
\tau_0 > \frac{1}{2 \eta}\bigg[\bigg( \frac{\epsilon x_{max}}{2} \bigg)^{2 \eta} \frac{\Gamma(1-\eta)}{\Gamma(1+\eta)}-1 \bigg].
\end{equation}
Note that as $\eta$ increases from 0 to $\sim0.1$, at which point the $\epsilon^{2\eta}$ term is highly suppressed relative to the $\epsilon^{0}$ term, this bound approaches $\tau_0 > -1/(2\eta)$, the previously obtained constraint for UV-brane localized fermion fields. If instead, $\eta$ decreases so that the fermions are localized near the TeV-brane, the lower bound in Eq.(\ref{t0GaugeLike}) becomes a negative number scaled by $\epsilon^{2 \eta}$, which in this regime should be large. This is consistent with the lack of constraints on $\tau_0$ in the TeV-brane localization scenario (\ie, if the only constraint on $\tau_0$ is that it must be greater than some very large negative value, then for practical purposes it has no constraints). It should also be noted that the bound in Eq.(\ref{t0GaugeLike}) is finite as $\eta \rightarrow 0$ (\ie, in the case of bulk gauge fields); in this limit the bound becomes
\begin{equation}
\tau_0 > \gamma +\log\bigg(\frac{\epsilon x_{max}}{2} \bigg),
\end{equation}
where here, $\gamma$ denotes Euler's constant.

\begin{figure}[htbp]
\centerline{\includegraphics[width=4.5in]{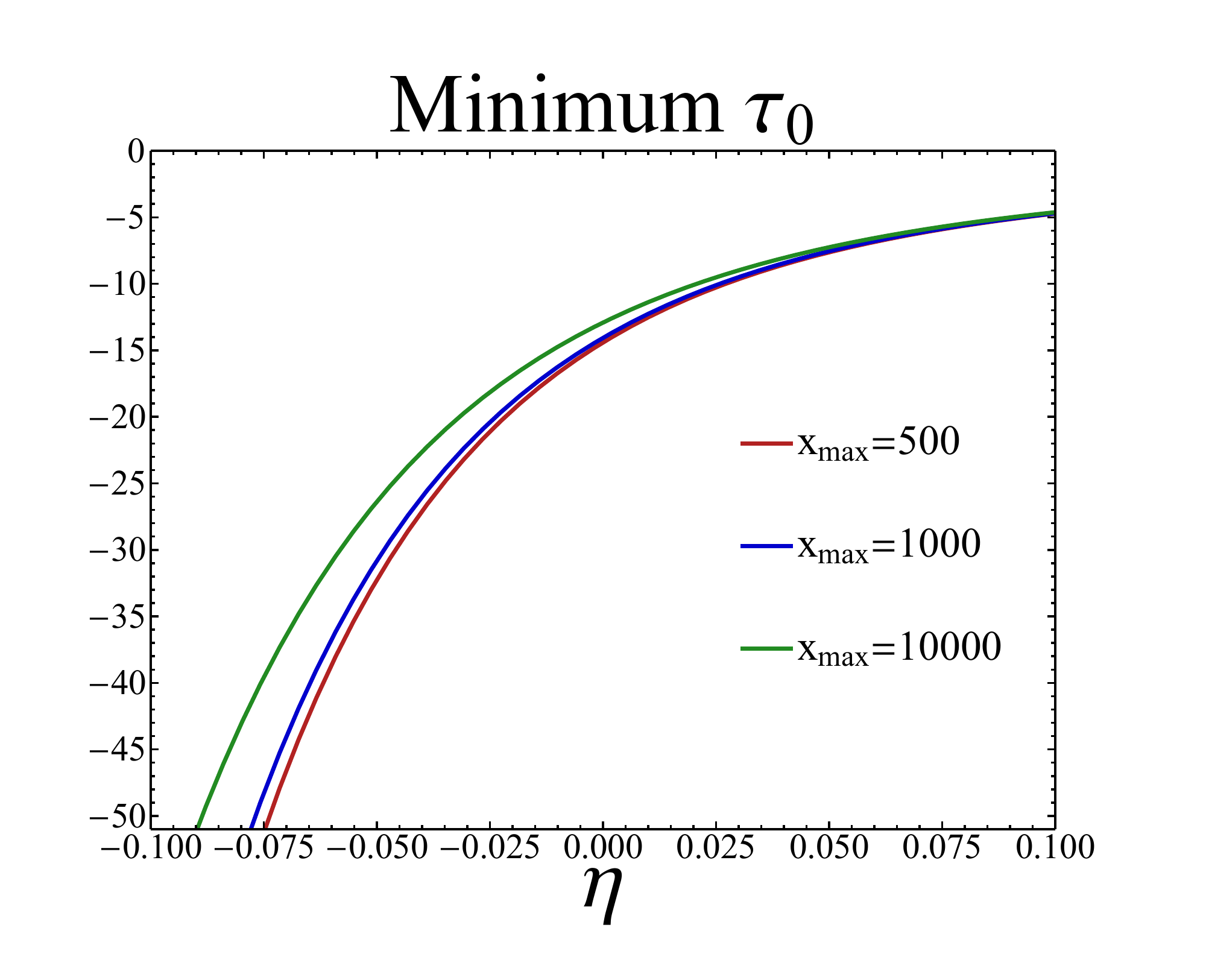}}
\vspace*{-0.10cm}
\caption{Lower bound on the parameter $\tau_0$ to avoid both ghosts and tachyonic roots below $x_{max}$, for $x_{max}=500$ (red), $1000$ (blue), or $10000$ (green).}
\label{figTachyonSafe}
\end{figure}

For numerical purposes, we should also acknowledge the possible dependence of the bound on $\tau_0$ on the specific choice of the cutoff, $x_{max}$. We see that in the region of interest, namely $-0.1 \lesssim \eta \lesssim 0.1$, the constraint on $\tau_0$ is only weakly dependent on the specific value of $x_{max}$; at worst, it is proportional to  $x_{max}^{\pm 0.2}$, due to the $x^{2 \eta}$ dependence depicted in Eq.(\ref{t0GaugeLike}) (when $\eta = 0$, the bound depends logarithmically on $x_{max}$). As a result, we see that a wide range of $x_{max}$ values produce essentially identical constraints. Numerically, we find the results shown in Fig \ref{figTachyonSafe} for the lower bound on $\tau_0$ as a function of $\eta$, for $x_{max}=500$, 1000, and 10000, to demonstrate the weak dependence of the boundaries on specific choices for $x_{max}$.

\subsubsection{The Case $\tau_\pi<0$}
We next consider the opposite situation, where $\tau_\pi < 0$. First, we explore the large-$x$ behavior of $f(x)$ in an attempt to eliminate some of this parameter space, based on the asymptotic expansion given in Eq.(\ref{UVCloseAsymptotic}). If a solution $f(x_{max})>0$ exists, then, by the same arguments given in the prior section, $f(x)>0$ for some $x< x_{max}$.

For $\tau_\pi < 0$, it is possible that the sign of the term $(1-(\frac{3}{8}+\eta+\frac{1}{2}\eta^2)\tau_\pi+x \tau_\pi)$ in Eq.(\ref{UVCloseAsymptotic}) is either positive or negative. However, in practice, for an $O(10^{2})$ or greater value for the cutoff $x_{max}$, $\tau_\pi$ would need simultaneously to be of order $O(10^{-2})$, or smaller, in order for $1+x \tau_\pi > 0$, which would be fine-tuned. Given that a natural value for the parameter $\tau_\pi$ is $\sim O(1-10)$ \cite{Davoudiasl:2002ua}, we find it unreasonable for $\tau_\pi$ to be small enough in magnitude to maintain $1+x \tau_\pi > 0$ for practical scenarios. Similarly, we do not explore the scenario where $1-(\frac{3}{8}+\eta+\frac{1}{2}\eta^2)\tau_\pi+x_{max}\tau_\pi=0$; because $x_{max}$ is an arbitrary cutoff parameter, any slight change in $x_{max}$ will eliminate this possibility. As a result, we will only consider the case where $1-(\frac{3}{8}+\eta+\frac{1}{2}\eta^2)\tau_\pi+x \tau_\pi < 0$, which yields the following condition on $\tau_0$ (employing the requirement that $f(x_{max})<0$ and the no-ghost condition)
\begin{equation}
\frac{(1-2 \eta \tau_\pi)\epsilon^{2 \eta}-1}{2 \eta} < \tau_0 < \frac{1}{2 \eta} \bigg[ \bigg( \frac{\epsilon x_{max}}{2} \bigg)^{2 \eta} \frac{\Gamma(1-\eta)}{\Gamma(1+\eta)} - 1 \bigg].
\end{equation}
Notably, the above condition also places a constraint on $\tau_\pi$. We can rewrite this condition as
\begin{equation}
    \frac{\epsilon^{2 \eta}-1}{2\eta}-\epsilon^{2 \eta} \tau_\pi < \frac{1}{2 \eta} \bigg[ \bigg( \frac{\epsilon x_{max}}{2} \bigg)^{2 \eta} \frac{\Gamma(1-\eta)}{\Gamma(1+\eta)} - 1 \bigg],
\end{equation}
and solving this inequality for $\tau_\pi$, we obtain
\begin{equation}
\tau_\pi > \frac{1}{2 \eta} \bigg[ 1-\bigg( \frac{x_{max}}{2} \bigg)^{2 \eta}\frac{\Gamma(1-\eta)}{\Gamma(1+\eta)}\bigg].
\end{equation}
The above lower bound on $\tau_\pi$ is negative for all $-0.1 \lesssim \eta \lesssim 0.1$ with a large cutoff $x_{max}$, so we still have a sizeable region of parameter space to probe for physical validity. To do so, we perform a numerical analysis. Using Mathematica \cite{mathematica}, a maximum of $f(x)$ in the region where $0<x<x_{max}$ is numerically determined at all points in this parameter space with natural brane terms ($|\tau_\pi|<50$ and $|\tau_0|<50$). To render the exploration of this parameter space tractable, the Taylor series expansion of Eq.(\ref{UVCloseTachyon}) is truncated at large $k$ and maximized, rather than attempting to maximize the exact function. Because of the overall $(k!)^{-2}$ suppression of each $(x/2)^{2k}$ term in this expansion, we find that keeping the first 500 terms of the Taylor series expansion is more than sufficient to estimate the value of $f(x)$ for $x<x_{max}=500$ with negligible error. Naively, the factor $C_k$ will only dominate the lower-order terms when $x>2k$, so even for $x_{max}=500$, the first 500 terms of the expansion are adequate for numerical purposes, just as in the case discussed in Sec \ref{UVSpecial} for UV-brane localized fermion fields. Searching for a region where all these conditions are satisfied produces a null set, indicating that the region $\tau_\pi < 0$ is disallowed by the existence of either ghost states or tachyonic Kaluza-Klein modes.

Summarizing, we find that the only region for a fermion field with localization close to $\eta=0$ that simultaneously avoids ghost states and tachyonic Kaluza-Klein modes is given by the conditions
\begin{equation}
\tau_\pi \geq 0,
\end{equation}
and
\begin{equation}
\tau_0 > \frac{1}{2 \eta}\bigg[\bigg( \frac{\epsilon x_{max}}{2} \bigg)^{2 \eta} \frac{\Gamma(1-\eta)}{\Gamma(1+\eta)}-1 \bigg].
\end{equation}
The latter, more difficult to visualize bound is depicted in Fig.\ref{figTachyonSafe}.

\subsection{Analysis: The Boundary Value Equation with Complex Masses}\label{ComplexMasses}

Thus far in this analysis, we have only addressed the possible existence of purely imaginary roots of Eq.(\ref{BoundaryEq}). However, in Sec. \ref{Framework} we asserted that the existence of $\textit{any}$ complex roots of Eq.(\ref{BoundaryEq}) would result in a phenomenologically unacceptable theory. We now address the possibility of general complex roots, and demonstrate that even if roots take on both real and imaginary non-zero parts, they will not result in any corresponding Kaluza-Klein particles in the 4-dimensional effective action. To begin, we note that a well-defined Kaluza-Klein state must be normalizable according to Eq.(\ref{Normalization}). This normalization condition is required in order to generate the equations of motion for the Kaluza-Klein states; for more detail see, \eg, Ref \cite{Davoudiasl:2000wi}. Using the definition of the bulk profiles $f_{L,R}$, we see that the normalization condition of a given Kaluza-Klein mode may be written in terms of the combination of Bessel functions $\zeta_{\frac{1}{2}-\nu}(z_n)$ as
\begin{equation}
\frac{1}{|N_n|^2} \int_{-\pi}^{\pi} d \phi e^{2 \sigma} \zeta_{\frac{1}{2}-\nu}(\frac{m_n}{k}e^\sigma) \zeta_{\frac{1}{2}-\nu}(\frac{m^*_n}{k}e^\sigma)(1+\Delta_{\tau_\pi,\tau_0})=1.
\end{equation}
Above, we have used the fact that $\nu$ is real, so that $\zeta^*_{\frac{1}{2}-\nu}(z) = \zeta_{\frac{1}{2}-\nu}(z^*)$.{\footnote{Recall that for the Bessel functions $J_{\eta}$ and $Y_{\eta}$, $(J_\eta (z))^*=J_{\eta^*}(z^*)$ and $(Y_\eta (z))^*=Y_{\eta^*}(z^*)$, for any $z$ not equal to a negative real number, and for all $\eta$.}} We have also again used the notation, $\Delta_{\tau_\pi,\tau_0} \equiv \frac{2}{kr_c}(\tau_\pi \delta(|\phi|-\pi)+\tau_0 \delta(\phi))$. Evaluating the above integral produces the result
\begin{align}
&\frac{2}{|N_n|^2 kr_c \epsilon^2} \bigg[ \frac{x_n^* \zeta_{-\frac{1}{2}-\nu}(x_n^*) \zeta_{\frac{1}{2}-\nu}(x_n)-x_n \zeta_{-\frac{1}{2}-\nu}(x_n) \zeta_{\frac{1}{2}-\nu}(x_n^*)}{(x_n)^2-(x_n^*)^2} + \tau_\pi \zeta_{\frac{1}{2}-\nu}(x_n^*) \zeta_{\frac{1}{2}-\nu}(x_n) \nonumber\\
\label{ComplexNormalization}\\
&-\frac{\epsilon x_n^* \zeta_{-\frac{1}{2}-\nu}(\epsilon x_n^*) \zeta_{\frac{1}{2}-\nu}(\epsilon x_n)-\epsilon x_n \zeta_{-\frac{1}{2}-\nu}(\epsilon x_n) \zeta_{\frac{1}{2}-\nu}(\epsilon x_n^*)}{(x_n)^2-(x_n^*)^2} +\epsilon^2 \tau_0 \zeta_{\frac{1}{2}-\nu}(\epsilon x_n^*) \zeta_{\frac{1}{2}-\nu}(\epsilon x_n)\bigg]=1. \nonumber
\end{align}
If $x_n$ is either purely real or purely imaginary, then the expression in the denominator $(x_n)^2-(x_n^*)^2$ becomes zero, and a limit must be taken to recover a meaningful expression (for both purely real and purely imaginary $x_n$, taking this limit yields a finite result for the above integral). However, if $x_n$ contains both real and imaginary parts, the above expression may be studied without the need to take any non-trivial limits. In this case, we may determine the normalization $N_n$ simply by inserting the standard boundary conditions below into Eq.(\ref{ComplexNormalization})
\begin{align}
\zeta_{-\frac{1}{2}-\nu}(x_n) &= x_n \tau_\pi \zeta_{\frac{1}{2}-\nu}(x_n),\\
\zeta_{-\frac{1}{2}-\nu}(\epsilon x_n) &= -\epsilon x_n \tau_0 \zeta_{\frac{1}{2}-\nu}(\epsilon x_n),\nonumber
\end{align}
which yields the following expression
\begin{align}
&\frac{2}{|N_n|^2 kr_c \epsilon^2} \bigg[ \bigg( \frac{(x_n^*)^2-(x_n)^2}{(x_n)^2-(x_n^*)^2}\bigg)\tau_\pi \zeta_{\frac{1}{2}-\nu}(x_n^*)\zeta_{\frac{1}{2}-\nu}(x_n)+\tau_\pi \zeta_{\frac{1}{2}-\nu}(x_n^*)\zeta_{\frac{1}{2}-\nu}(x_n)\nonumber\\
\\
&-\bigg( \frac{(x_n)^2-(x_n^*)^2}{(x_n)^2-(x_n^*)^2}\bigg)\epsilon^2 \tau_0 \zeta_{\frac{1}{2}-\nu}(x_n^*)\zeta_{\frac{1}{2}-\nu}(x_n)+\epsilon^2 \tau_0 \zeta_{\frac{1}{2}-\nu}(\epsilon x_n^*)\zeta_{\frac{1}{2}-\nu}(\epsilon x_n)\bigg]=0.\nonumber
\end{align}
Thus, the bulk wave functions of a complex-mass fermionic Kaluza-Klein mode will be ``orthogonal to themselves", implying that these wave functions are unphysical, \ie, impossible to normalize. As a result, we find that even if roots of Eq.(\ref{BoundaryEq}) with nonzero real and imaginary parts exist, they will not, in fact, produce normalizable Kaluza-Klein states. Furthermore, we see that these bulk fields would $\textit{vanish}$ from the Lagrangian after integration over $\phi$. Thus, we find that the only possible physical particles arising in the case of fermion fields with generic BLKT's will have either purely real or purely imaginary masses.

\section{Presence of Spontaneous Symmetry Breaking}\label{SSBDiscussion}

Thus far, we have adopted the simplifying assumption that the fermion fields we consider are not subject to any form of spontaneous symmetry breaking (SSB). However, except for the possibility of neutrinos, all fermion fields in the Standard Model acquire mass via the conventional Higgs mechanism. In the Randall-Sundrum framework, the Higgs field is generally localized on the TeV-brane, in order to effect a hierarchy between the weak scale (set by the 4-dimensional Higgs vev) and the Planck scale. In this section, we discuss the effects of adding SSB as a perturbation, and demonstrate that it is unlikely to alter the conclusions we have arrived at above. In particular, we probe the possibility of SSB eliminating through the Higgs mechanism the tachyons or ghost states that are present in the theory; given the fact that the majority of parameter space for this model is eliminated by our analysis above in the absence of SSB, this question is of no small importance. In particular, we shall demonstrate that in most regions of parameter space, the modifications to the constraints on brane terms and localizations required to prevent ghost states for the lowest-lying KK tower modes (corresponding to the SM particles) in the absence of SSB will be very small. Further, we will show that rather than helping to eliminate a tachyonic root that might arise in the case without SSB, the presence of the Higgs mechanism will to first approximation merely move an existing tachyonic root along the imaginary line, and to higher order (at best) move it slightly into the general complex plane. In short, if a point in parameter space is disallowed in a theory without the Higgs mechanism, we shall demonstrate that it is very likely still disallowed when the Higgs vev is introduced.

To begin, we must introduce a set of Yukawa couplings into the theory. The simplest fermion action with Yukawa couplings necessarily involves two bulk fermion fields, denoted here by $Q$ and $q$, that in the absence of SSB will produce a left-handed zero-mode and a right-handed zero-mode, respectively. The Higgs mechanism then mixes these fields and produces a single massive fermionic field out of the two chiral zero-mode states, as well as altering the spectrum of the members of both KK towers. This action may be written, analogously to Eq.(\ref{fermionAction}) as
\begin{eqnarray}\label{SSBAction}
 S_F= &\frac{}{} &\int d^4 x \int r_c d\phi \, \sqrt{G} \, 
\left\{V^M_N (\frac{i}{2}{\overline Q} \Gamma^N \partial_M Q+\frac{i}{2}{\overline q} \Gamma^N \partial_M q +h.c.) \right. \nonumber\\
& + &
\left. [2\tau_0/kr_c ~\, \delta(\phi) + 2\tau_\pi/kr_c ~\, \delta(|\phi| - \pi)] \,V^\mu_\nu (i{\overline Q}_L \gamma^\nu \partial_\mu Q_L +i{\overline q}_R \gamma^\nu \partial_\mu q_R+h.c.) \right. \nonumber\\
& - &
\left. sgn(\phi) k(\nu_Q {\overline Q} q-\nu_q {\overline q} q )\frac{}{} \right.\\
& - &
\left. \frac{2}{kr_c}\delta(|\phi|-\pi)e^{\sigma} \frac{v}{\sqrt{2} }[{\overline Q}_L Y q_R+{\overline q}_R Y^* Q_L] \frac{}{} \right\} \nonumber\,.
\end{eqnarray}
Here, $v$ denotes the 4-dimensional Higgs vacuum expectation value, $v \sim 246 \GeV$. Following \cite{casagrande}, the Yukawa coupling $Y$ is taken to be $O(1)$, and of arbitrary complex phase. Note that here, we assume for simplicity that both $Q$ and $q$ have identical BLKT's $\tau_\pi$ and $\tau_0$. Given that brane terms must likely be all approximately the same order of magnitude $\sim O(1-10)$ to be natural, it is not unreasonable to expect that the general case of both fields having independent brane terms will be qualitatively similar to the case where the brane terms are universal. 

As in the case without SSB, we want a 4-dimensional action of the form,
\begin{equation}\label{GoalAction}
\sum_{n} \int d^4 x \, \big[ \bar{f}^{(n)} i \slashed{\partial} f^{(n)}-m_n \bar{f}^{(n)} f^{(n)} \big].
\end{equation}
Here, we note that the summation extending over the Kaluza-Klein modes is defined differently in this scenario than it is for a single fermion bulk field. In the absence of SSB, a single bulk field would have a massless $Z_2$-even zero-mode, and an infinite tower of pairs of Kaluza-Klein fermion fields, one $Z_2$-even and the other $Z_2$-odd. In the presence of SSB, however, the Yukawa term in the action mixes the $\textit{two}$ bulk fermion fields. The result is that the index $n$ extends over $\textit{twice}$ as many KK tower modes, all of which are now admixtures of $Z_2-$even and $Z_2-$odd bulk wave functions (in particular, the left-handed KK modes $f^{(n)}_L$ will be mixtures of the $Q$ field's $Z_2-$even modes and $q$ field's $Z_2-$odd modes, while the right handed modes $f^{(n)}_R$ will be mixtures of the $Q$ field's $Z_2-$odd modes and $q$ field's $Z_2-$even modes). While it is reasonable to think of all of these states as simple perturbations of the separate KK towers for the $Q$ and $q$ fields, in general it is difficult to associate a given mode here to a perturbation of a corresponding mode in the absence of the Higgs mechanism. As a result, we adopt the simplistic index $n$, understanding that the summation now extends over the expanded set of mixed states.

To begin, we perform Kaluza-Klein decompositions on $Q$ and $q$ in a similar fashion as given in Eq.(\ref{PsiLR}). As noted above, the $Q$ field's $Z_2-$even modes are left-handed, while its $Z_2-$odd modes are right-handed, while the $q$ field's modes have the opposite chirality. Following the notation of \cite{casagrande}, we refer to $Z_2-$even bulk profiles for the $n^{th}$ mode of the $Q(q)$ fields as $C^{(Q(q))}_n(\phi)$, and the corresponding $Z_2-$odd bulk profiles as $S^{(Q(q))}_n(\phi)$. Including this notation in our Kaluza-Klein decomposition yields the following expansions
\begin{align}\label{SSBKK}
Q_L = \sum_{n}\frac{e^{2 \sigma}}{\sqrt{r_c}}C^{(Q)}_n(\phi) f_L^{(n)}(x), && Q_R = \sum_{n}\frac{e^{2 \sigma}}{\sqrt{r_c}}S^{(Q)}_n(\phi) f_R^{(n)}(x),\\
q_L = \sum_{n}\frac{e^{2 \sigma}}{\sqrt{r_c}}S^{(q)}_n(\phi) f_L^{(n)}(x), && q_R = \sum_{n}\frac{e^{2 \sigma}}{\sqrt{r_c}}C^{(q)}_n(\phi) f_R^{(n)}(x). \nonumber
\end{align}
Inserting these expansions in the action given by Eq.(\ref{SSBAction}) yields the following conditions for canonically normalized kinetic terms in the effective 4-dimensional action (compare with Eq.(\ref{NonSSBNorm}))
\begin{align}
    \int_{-\pi}^{\pi} d \phi e^{\sigma} \left[ C_m^{(Q)*}(\phi) C_n^{(Q)}(\phi)(1+\Delta_{\tau_\pi,\tau_0})+S_m^{(q)*}(\phi) S_n^{(q)}(\phi) \right] &= \delta_{m n}, \label{SSBNormL1}\\
    \nonumber \\
    \int_{-\pi}^{\pi} d \phi e^{\sigma} \left[ C_m^{(q)*}(\phi) C_n^{(q)}(\phi)(1+\Delta_{\tau_\pi,\tau_0})+S_m^{(Q)*}(\phi) S_n^{(Q)}(\phi) \right] &= \delta_{m n}. \nonumber
\end{align}
In order to produce the mass term of Eq.(\ref{GoalAction}), we require, in analogy to Eq.(\ref{NonSSBR}), that the bulk wave functions satisfy the equations of motion
\begin{align}
    \left( \frac{1}{r_c} \partial_\phi +sgn(\phi) \nu_Q k \right) S_n^{(Q)} &= m_n (1+\Delta_{\tau_\pi,\tau_0})e^\sigma C_n^{(Q)}-e^\sigma \delta(|\phi|-\pi)\frac{\sqrt{2}v Y}{k r_c}C_n^{(q)}, \nonumber\\
    \left( \frac{1}{r_c} \partial_\phi - sgn(\phi) \nu_Q k \right) C_n^{(Q)} &= -m_n e^{\sigma} S_n^{(Q)},\\
    \left( \frac{1}{r_c} \partial_\phi + sgn(\phi) \nu_q k \right) S_n^{(q)} &= -m_n (1+\Delta_{\tau_\pi,\tau_0})e^\sigma C_n^{(q)}+e^\sigma \delta(|\phi|-\pi)\frac{\sqrt{2}v Y^*}{k r_c}C_n^{(Q)}, \nonumber\\
    \left( \frac{1}{r_c} \partial_\phi -sgn(\phi) \nu_q k \right) C_n^{(q)} &= m_n e^{\sigma} S_n^{(q)}. \nonumber
\end{align}
Notably, with the exception of the additional boundary terms proportional to the Higgs vev $v$, which only appear on the brane and therefore affect only boundary conditions, the differential equations for the $Q$ and $q$ fields are identical to Eq.(\ref{NonSSBR}). So, in analogy with the case neglecting SSB, the general solutions of these equations of motion are
\begin{align}
    C_n^{(Q)}(\phi) &=\frac{e^{\sigma/2}}{N^Q_n}\zeta_{1+\eta_Q}(z_n), && S_n^{(Q)}(\phi) =\frac{-sgn(\phi)e^{\sigma/2}}{N^Q_n}\zeta_{\eta_Q}(z_n), \label{CSq}\\
    C_n^{(q)}(\phi) &=\frac{e^{\sigma/2}}{N^q_n}\zeta_{1+\eta_q}(z_n), && S_n^{(q)}(\phi) =\frac{sgn(\phi)e^{\sigma/2}}{N^q_n}\zeta_{\eta_q}(z_n),\nonumber
\end{align}
where $\eta_{Q,q}$ is defined in analogy to our treatment of the case without SSB (\ie, $\eta_{Q,q} \equiv -\frac{1}{2}-\nu_{Q,q}$), while the $\zeta$ functions are defined by Eq.(\ref{zetaDef}). Note that since the UV-brane ($\phi=0$) boundary conditions in this setup are equivalent to those in the absence of the Higgs, the constants $\alpha_n$ and $\beta_n$ in the definition of $\zeta_{1+\eta_q}(z_n)$ are still given by Eq.(\ref{alphan}). Inserting these expressions for the bulk profiles into Eq.(\ref{SSBNormL1}) yields the following coupled expressions for $N^Q_n$ and $N^q_n$
\begin{align}
1 &= \int_{-\pi}^{\pi} d \phi \left\{ \frac{e^{2 \sigma}}{|N^Q_n|^2}(1+\Delta_{\tau_\pi,\tau_0})|\zeta_{1+\eta_Q}(z_n)|^2 + \frac{e^{2 \sigma}}{|N^q_n|^2}|\zeta_{\eta_q}(z_n)|^2 \frac{}{} \right\}, \label{SSBNormQ}\\
1 &= \int_{-\pi}^{\pi} d \phi \left\{ \frac{e^{2 \sigma}}{|N^q_n|^2}(1+\Delta_{\tau_\pi,\tau_0})|\zeta_{1+\eta_q}(z_n)|^2 + \frac{e^{2 \sigma}}{|N^Q_n|^2}|\zeta_{\eta_Q}(z_n)|^2 \frac{}{} \right\}. \nonumber
\end{align}

The introduction of the additional SSB terms on the TeV-brane results in significant modifications to the TeV-brane boundary conditions, which govern the spectrum of states in the effective four-dimensional theory. The TeV-brane boundary conditions now become (compare with Eq.(\ref{basicBoundary}))
\begin{align}
\zeta_{\eta_q}(x_n)-x_n \tau_\pi \zeta_{1+\eta_q}(x_n) &=-\frac{v Y^*}{\sqrt{2} M_{KK}}\bigg( \frac{N^q_n}{N^Q_n} \bigg) \zeta_{1+\eta_Q}(x_n), \label{SSBQ}\\
\zeta_{\eta_Q}(x_n)-x_n \tau_\pi \zeta_{1+\eta_Q}(x_n) &=-\frac{v Y}{\sqrt{2} M_{KK}}\bigg( \frac{N^Q_n}{N^q_n} \bigg) \zeta_{1+\eta_q}(x_n). \nonumber
\end{align}
$N^{Q(q)}_n$ refers to the normalization of the $Q$ $(q)$ wave function, selected to produce an action of the form of Eq.(\ref{GoalAction}). It is interesting to note that $\zeta_{1+\eta_{q,Q}}(x_n)$ approaches 0 as $x_n$ becomes very large; as a result, while the addition of SSB can have a significant effect on low-lying KK modes (in particular, the massless chiral zero-modes become massive SM fermions), the more massive tower states should be significantly less affected by SSB. Multiplying the top and bottom equations in (\ref{SSBQ}) together, one arrives at an equation for the mass spectrum that eliminates any dependence on the normalization factors $N^Q_n$ and $N^q_n$,
\begin{equation}\label{SSBBoundary}
[\zeta_{\eta_q}(x_n)-x_n \tau_\pi \zeta_{1+\eta_q}(x_n)][\zeta_{\eta_Q}(x_n)-x_n \tau_\pi \zeta_{1+\eta_Q}] = \frac{v^2 |Y|^2}{2 M_{KK}^2} \zeta_{1+\eta_Q}(x_n) \zeta_{1+\eta_q}(x_n).
\end{equation}
Armed with these equations, then, it is in principle possible, as in the case without SSB, to derive the wave functions and masses of the entire KK tower with the full inclusion of the effects of SSB. In the following sections we explore the effects of SSB on points in the parameter space that, in the absence of these effects, are disallowed by the existence of tachyonic KK modes or ghost states.

\subsection{Ghost States in the Presence of SSB}\label{SSBNormSec}

Using the framework discussed above, we now derive the conditions for avoiding ghost states equivalent to those discussed in the previous case without SSB; namely, in what cases are the normalizations of the lowest-lying KK mode ghost-like (that is, $|N^Q_0|^2<0$ or $|N^q_0|^2<0$). Restricting our analysis to the perturbative regime, where $v^2/M_{KK}^2 \ll 1$ is assumed (which corresponds to physical expectations), we begin by determining the location of the root in Eq.(\ref{SSBBoundary}). Expanding Eq.(\ref{SSBBoundary}) to the lowest order in $x^2$, we arrive at the following result for the lowest-lying root
\begin{equation}\label{approxX}
x_0^2 = \frac{m_0^2}{M_{KK}^2} \bigg[ 1+\frac{1}{4}
\frac{m_0^2}{M_{KK}^2}\bigg( \frac{1}{\eta_q}\bigg( 1-\frac{\epsilon^{-2 \eta_q}(1+2 \eta_q \tau_0)}{1+\eta_q}\bigg)+\frac{1}{\eta_Q}\bigg( 1-\frac{\epsilon^{-2\eta_Q}(1+2 \eta_Q \tau_0)}{1+\eta_Q}\bigg)\bigg) \bigg].
\end{equation}
Here, $m_0^2$ is given by the expression
\begin{equation}
m_0^2 \equiv \frac{v^2 |Y|^2}{2}\bigg( \frac{4 \eta_Q \eta_q}{[(1-2 \eta_Q \tau_\pi)-\epsilon^{-2 \eta_Q}(1+2 \eta_Q \tau_0)][(1-2 \eta_q \tau_\pi)-\epsilon^{-2 \eta_q}(1+2 \eta_q \tau_0)]}\bigg).
\end{equation}
Notably, $m_0^2$ is precisely the mass arising from the Yukawa coupling that the particle formed from the two individual chiral zero-modes (with a bulk profile of Eq.(\ref{fnL0})) of $Q$ and $q$ $\textit{would have}$, in the absence of any mixing with additional KK tower modes. It is also notable that $m_0^2$ is proportional to the product of the normalizations of both of these zero modes. Hence, we see a connection between the no-ghost condition of the case where SSB is neglected, given by Eq.(\ref{Ghosts}), and the restriction on the allowable parameter space of the case with SSB: If either one of $Q_L^{(0)}$ or $q_R^{(0)}$ fails to satisfy Eq.(\ref{Ghosts}), then $m_0^2<0$. Since $m_0^2$ is proportional to the Yukawa-induced mass squared of the lowest-lying KK mode up to $O(\frac{m_0^2}{M_{KK}^2})$ corrections, this would indicate that this lowest-lying mode, rather than serving its purpose as a massive SM fermion, would then be a particle of some imaginary (tachyonic) mass. Naturally, this is phenomenologically unacceptable. The one exception to this conclusion, however, would be the scenario where $\textit{both}$ the $Q_L^{(0)}$ and $q_R^{(0)}$ states would be ghost-like in the absence of the Higgs mechanism. In this case, $m_0^2$ would be positive, and so naive analysis would suggest a physical mass for the lowest-lying KK mode when SSB is applied. Hence, the scenario where either the $Q$ or $q$ field possesses a ghost-like zero-mode when SSB is neglected, but the other does not, is easily dismissed as unphysical. However, the scenario where both $Q$ and $q$ produce ghost-like zero modes when the Higgs field is ignored still produces a positive $m_0^2$, and hence requires further inspection.

To continue exploring the conditions under which ghost-like states are produced in the presence of SSB, we consider the normalization condition of Eq.(\ref{SSBNormQ}). Performing the integration for some real $m_n$ yields (after applying the UV-brane boundary condition $\zeta_{\eta_{Q,q}}(\epsilon x_n)=-\epsilon x_n \tau_0 \zeta_{1+\eta_{Q,q}}(\epsilon x_n)$)
\begin{equation}\label{RealNorm}
\begin{multlined}
\frac{1}{|N^Q_n|^2} [x_n^2 (1+2 \tau_\pi)\zeta_{1+\eta_Q}^2(x_n)-2 x_n (1+\eta_Q)\zeta_{\eta_Q}(x_n)\zeta_{1+\eta_Q}(x_n)+x_n^2 \zeta_{\eta_Q}^2(x_n)\\
-\epsilon^2 x_n^2(1+2 \eta_Q \tau_0+\epsilon^2 x_n^2 \tau_0^2) \zeta_{1+\eta_Q}^2(\epsilon x_n)]\\
+\frac{1}{|N^q_n|^2} [x_n^2 \zeta_{1+\eta_q}^2(x_n)-2 x_n \eta_q \zeta_{\eta_q}(x_n)\zeta_{1+\eta_q}(x_n)+x_n^2 \zeta_{\eta_q}^2(x_n)\\
-\epsilon^2 x_n^2(1+2 \eta_q \tau_0+\epsilon^2 x_n^2 \tau_0^2) \zeta_{1+\eta_q}^2(\epsilon x_n)]= kr_c \epsilon^2 x_n^2.
\end{multlined}
\end{equation}
We may now address the normalization of the lowest-lying mode by expanding this expression around $x_0 \approx 0$. Up to $O(x_0^2)$, Eq.(\ref{RealNorm}) may be approximated as
\begin{equation}\label{NormApprox}
\begin{multlined}
\pi^2 kr_c \epsilon^2 x_0^2 |N_0^Q|^2 \approx 4\frac{[(1+2 \eta_Q \tau_0)-(1-2 \eta_Q \tau_\pi)\epsilon^{2 \eta_Q}]}{\eta_Q}\\
+2 x_0^2 \bigg[ \frac{\epsilon^{2 \eta_Q}}{(1-\eta_Q)\eta_Q}+\frac{2 \tau_\pi \epsilon^{2 \eta_Q}}{(1+\eta_Q)\eta_Q}-\frac{(1+2 \tau_\pi)(1+2 \eta_Q \tau_0)}{(1+\eta_Q)\eta_Q} \bigg]\\
+\frac{|N_0^Q|^2}{|N_0^q|^2} x_0^2 \bigg[ \frac{\epsilon^{2 \eta_q}}{(1-\eta_q) \eta_q^2}-\frac{2(1+2\eta_q \tau_0)}{\eta_q^2}+\frac{\epsilon^{-2 \eta_q}(1+2 \eta_q \tau_0)^2}{(1+\eta_q)\eta_q^2}\bigg].
\end{multlined}
\end{equation}
From this, we arrive at an expression for the ratio of the normalizations $|N^Q|^2/|N^q|^2$, using Eq.(\ref{SSBQ}). Specifically, by dividing the bottom expression in Eq.(\ref{SSBQ}) by the conjugate of the top expression, we arrive at the following leading order expression
\begin{align}
\frac{|N^Q|^2}{|N^q|^2} &= \frac{(\zeta_{\eta_Q}(x_0)-x_0 \tau_\pi \zeta_{1+\eta_Q}(x_0))\zeta_{1+\eta_Q}(x_0)}{(\zeta_{\eta_q}(x_0)-x_0 \tau_\pi \zeta_{1+\eta_q}(x_0))\zeta_{1+\eta_q}(x_0)} \\
&\approx \frac{\eta_q}{\eta_Q} \frac{[(1+2 \eta_Q \tau_0)-\epsilon^{2 \eta_Q}(1-2 \eta_Q \tau_\pi)]}{[(1+2 \eta_q \tau_0)-\epsilon^{2 \eta_q}(1-2 \eta_q \tau_\pi)]} + O(x_0^2). \nonumber
\end{align}
Note that because the only term proportional to this ratio in Eq.(\ref{NormApprox}) is already proportional to $x_0^2$, we only need to keep the $x_0^0$ term above for our purposes. Finally, we insert the expression for $x_0^2$ given in Eq.(\ref{approxX}) (dropping the higher order terms of $O(v^4/M_{KK}^4)$) to arrive at the following expression for our normalization condition
%
%
\begin{equation}\label{NormApprox2}
    kr_c \epsilon^2 x^2 |N^Q|^2 \approx \frac{4}{\pi^2}\frac{1}{\lambda_Q} \bigg\{ 1+\frac{v^2 |Y|^2}{2 M_{KK}^2}\epsilon^{2(\eta_Q+\eta_q)}\lambda_Q \lambda_q \bigg[ \lambda_Q \rho_Q+\lambda_q \xi_q \bigg] \bigg\},
\end{equation}
where we have defined

\begin{align}
    \lambda_{Q,q} &\equiv \frac{\eta_{Q,q}}{[(1+2 \eta_Q \tau_0)-(1-2 \eta_{Q,q} \tau_\pi)\epsilon^{2 \eta_{Q,q}}]}, \nonumber\\
    \rho_Q &\equiv \bigg( \frac{2 \epsilon^{2 \eta_Q}}{(1-\eta_Q)\eta_Q}+\frac{4 \tau_\pi \epsilon^{2 \eta_Q}}{\eta_Q}-\frac{2(1+2 \tau_\pi)(1+2 \eta_Q \tau_0)}{(1+\eta_Q)\eta_Q}\bigg),\\
    \xi_q &\equiv \bigg( \frac{\epsilon^{2 \eta_q}}{(1-\eta_q)\eta_q^2}-\frac{2(1+2 \eta_q \tau_0)}{\eta_q^2}+\frac{\epsilon^{-2 \eta_q}(1+ 2\eta_q \tau_0)^2}{(1+\eta_q)\eta_q^2}\bigg).\nonumber
\end{align}
In a similar fashion, an analogous expression may be derived for $|N^q|^2$, with the only difference being the interchange of the $Q$ and $q$ sub- and superscripts in the above expression. Notably, if the $O(v^2/M_{KK}^2)$ corrections are neglected, both $|N^Q|^2$ and $|N^q|^2$ will yield negative norm squared values only when the condition of Eq.(\ref{Ghosts}) is violated for a specific fermionic field (\ie, a violation for $Q$ will yield a ghost-like $Q$ state, and a violation for $q$ will yield a ghost-like $q$ state). 
A detailed inspection of this correction term's behavior throughout the full parameter space is beyond the scope of this paper, but several observations can be made. Notably, if $\eta_Q$ is allowed to be large and positive enough to render the $\epsilon^{2 \eta_Q}$ terms insignificant (for consistency with our prior analysis of the case without SSB, this may be assumed to be approximately at $\eta_Q \gsim 0.1$), the leading correction terms for the normalization factors become suppressed by $\epsilon^{2 \eta_Q}$. This suggests that in order to make these correction terms large enough to flip the sign of the normalization, $v^2/M_{KK}^2$ would likely have to be extremely large, directly counter to our assumption that $v^2/M_{KK}^2 \ll 1$. The other limit, in which both fermions are localized near the TeV-brane (and hence $\eta_{Q,q} \lesssim -0.1$), presents more interesting behavior. In this case, the magnitude of the $v^2/M_{KK}^2$ correction term may be essentially arbitrarily increased by tuning $\tau_\pi$ and $\eta_{Q,q}$ such that $1-2\eta_{Q,q} \tau_\pi \approx 0$. In the event of $Q$ and $q$ both violating the previous condition for avoiding ghosts, Eq.(\ref{Ghosts}), this in fact results in a $\textit{negative}$ proportional correction to both $|N^Q|^2$ and $|N^q|^2$ of arbitrary magnitude, suggesting that it is in fact possible, in principle, in carefully tuned regions of parameter space for a model to lack ghosts when spontaneous symmetry breaking effects are included, while possessing them in the absence of SSB. However, in practice, tuning $(1-2 \eta_{Q,q} \tau_\pi) \approx 0$ also arbitrarily increases the value of $x_0^2$ (from Eq.(\ref{approxX})), which runs the risk of rendering the perturbative approximation for the normalization invalid. Furthermore, the $O(v^2/M_{KK}^2)$ correction terms to $x_0^2$ that were dropped in Eq.(\ref{NormApprox2}) would also be arbitrarily increased, rendering the results in this region that employed any perturbative calculations suspect. In fact, a cursory examination of the general case given in Eq.(\ref{NormApprox2}) suggests a similar conclusion for the entire parameter space: The only method to increase the correction terms to $|N^{Q,q}|^2$ arbitrarily, as would be necessary to alter their sign, would require a degree of tuning to achieve $(1+2 \eta_{Q,q} \tau_0)-(1-2 \eta_{Q,q} \tau_\pi)\epsilon^{2 \eta_Q} \approx 0$, which will in turn result in an arbitrary increase in the value of $x_0^2$, and this value of $x_0^2$ may deviate significantly from the $O(v^2/M_{KK}^2)$ approximation employed in Eq.(\ref{NormApprox2}). As a result, we close our discussion on the possibility of rendering states that violate Eq.(\ref{Ghosts}) phenomenologically viable via the presence of spontaneous symmetry breaking; we find that at best it only may be possible in finely tuned regions of parameter space, and these regions rapidly fall out of the range of validity of the perturbative approximation employed here.

\subsection{Tachyonic Roots in the Presence of SSB}

Having explored the scenario where the effects of SSB may eliminate the presence of ghosts in the RS model, we now move on to addressing the effect of SSB on tachyonic roots that appear in our analysis without SSB. First, we note that the existence of spontaneous symmetry breaking should not make a given root of Eq.(\ref{SSBBoundary}) disappear altogether; as we have noted in previous sections, the expressions for $\zeta$ employed here are well-approximated by a truncated polynomial series, where we use the identities
\begin{align}\label{ZetaIDs}
    \zeta_{\eta}(x) &= \frac{1}{\eta \pi} \sum_{k=0}^{\infty} \bigg( \frac{x}{2} \bigg)^{2 k} \frac{(-1)^k}{k!}\bigg[ \frac{\Gamma(1-\eta)\epsilon^{\eta}}{\Gamma(1+k-\eta)}-\frac{(1+2 \eta \tau_0)\Gamma(1+\eta)\epsilon^{-\eta}}{\Gamma(1+k+\eta)}\bigg],\\
    \zeta_{1+\eta}(x) &= \frac{-2}{\eta \pi x} \sum_{k=0}^{\infty} \bigg( \frac{x}{2} \bigg)^{2 k} \frac{(-1)^k}{k!}\bigg[ \frac{(k-\eta)\Gamma(1-\eta)\epsilon^{\eta}}{\Gamma(1+k-\eta)}-\frac{k(1+2 \eta \tau_0)\Gamma(1+\eta)\epsilon^{-\eta}}{\Gamma(1+k+\eta)}\bigg].\nonumber
\end{align}
Inserting these expressions into Eq.(\ref{SSBBoundary}), the product $\zeta_{1+\eta_Q}(x)\zeta_{1+\eta_q}(x)$ introduces a single $x^{-2}$ term into Eq.(\ref{SSBBoundary}) when $v \neq 0$ which is not present when $v = 0$. By applying the fundamental theorem of algebra to truncated versions of the series expressions for $\zeta_{\eta}(x)$ and $\zeta_{1+\eta}(x)$, we expect that Eq.(\ref{SSBBoundary}) should have two additional roots when $v\neq 0$ (corresponding to $\pm$ of the mass of the SM particle in the model), in comparison to the case with $v=0$. We note that this is only true when counting the multiplicities of roots of the equations, \eg, if a given value of $x$ is a double root of Eq.(\ref{SSBBoundary}) when $v=0$, we can expect this root to be split into two when $v\neq 0$. This will be discussed in greater detail below. As noted in Sec \ref{SSBNormSec}, to lowest order in $v^2/M_{KK}^2$, the roots corresponding to the SM particle will be real whenever the non-SSB no-ghost condition (Eq.(\ref{Ghosts})) is satisfied for both the $Q$ and $q$ fields. As we have already determined that scenarios where these conditions are violated are unlikely to produce physical models, the possibility of the additional roots introduced by SSB being tachyonic or complex will not be addressed further here. Therefore, the problem of determining if SSB can ``save" a region of parameter space that is disallowed in its absence can be reduced to determining how the existing tachyonic roots of Eq.(\ref{SSBBoundary}) are modified when $v \neq 0$. If imaginary roots can be rendered real, then the presence of SSB will open new regions of allowed parameter space.

Again, we will work in the regime where the SSB terms in Eq.(\ref{SSBBoundary}) represent a small perturbation, and we first determine the lowest-order (in $v^2/M_{KK}^2$) correction to the location of a root, $x_0$, of Eq.(\ref{SSBBoundary}) when $v=0$. To begin, we expand Eq.(\ref{SSBBoundary}) about $x_0$, assuming (without loss of generality) that $x_0$ would correspond to a $KK$ tower mode of the $Q$, rather than $q$, field in the absence of SSB. Noting that $\zeta_{\eta_Q}(x_0)=x_0 \tau_\pi \zeta_{1+\eta_Q}(x_0)$, this yields
\begin{multline}\label{FirstOrderSoln}
(x'_0-x_0)[\frac{d}{dx}Z_{\eta_Q}(x)|_{x=x_0}][Z_{\eta_q}(x_0)]\\
\approx \frac{v^2 |Y|^2}{2 M_{KK}^2}(\zeta_{1+\eta_Q}(x_0)\zeta_{1+\eta_q}(x_0)+(x'_0-x_0)\frac{d}{dx}(\zeta_{1+\eta_Q}(x)\zeta_{1+\eta_q}(x))|_{x=x_0}).
\end{multline}
Here, we refer to the perturbed position of the root in the presence of SSB as $x'_0$, and we have defined
\begin{equation}
    Z_{\eta}(x) \equiv \zeta_{\eta}(x)-x \tau_\pi \zeta_{1+\eta}(x).
\end{equation}
Eq.(\ref{FirstOrderSoln}) can be solved for the shift in the root, $(x'_0-x_0)$, and yields the result
\begin{equation}\label{SSBLowestOrder}
    (x'_0-x_0) \approx \frac{v^2 |Y|^2}{2 M_{KK}^2} \bigg(\frac{\zeta_{1+\eta_Q}(x_0) \zeta_{1+ \eta_q}(x_0)}{Z_{\eta_q}(x_0)\frac{d}{dx}Z_{\eta_Q}(x)|_{x=x_0})-\frac{v^2 |Y|^2}{2 M_{KK}^2}\frac{d}{dx}(\zeta_{1+\eta_Q}(x)\zeta_{1+\eta_q}(x))|_{x=x_0}}\bigg).
\end{equation}
To help shed some light on the implications of the lowest-order correction to $x_0$, we now employ the power series identities in Eq.(\ref{ZetaIDs}). These expressions in turn prove illuminating for the complex phases of $\zeta_{\eta}(x)$, $\zeta_{1+\eta}(x)$, and their derivatives when $x$ is purely imaginary or purely real. Since the gamma functions, exponentials, and factorials that appear in the expressions for $\zeta_{\eta}(x)$ and $\zeta_{1+\eta}(x)$ in Eq.(\ref{ZetaIDs}) are real, any complex phase of these functions must arise from a complex phase of $x$ itself. If $x$ is purely imaginary, then, any even power of $x$ will be real, while any odd power of $x$ will be imaginary. Therefore, $\zeta_{\eta}(x)$ is real for purely imaginary $x$, because $\zeta_{\eta}(x)$ contains only even powers of $x$, while $\zeta_{1+\eta}(x)$ is imaginary for purely imaginary $x$, because it contains only odd powers of $x$. The same logic can easily be applied to the expression $Z_{\eta}(x)=\zeta_{\eta}(x)-x \tau_\pi \zeta_{1+\eta}(x)$: Since it contains only even powers of $x$, it is real when $x$ is imaginary. 

Expanding this argument to include the derivatives of these $\zeta$ functions when $x$ is imaginary is straightforward, since each derivative with respect to $x$ turns a term with an odd power of $x$ into one with an even power, and vice versa. Thus, $\frac{d}{dx}(\zeta_{\eta}(x)-x \tau_\pi \zeta_{1+\eta}(x))$ has only odd powers of $x$, and is hence imaginary when $x$ is imaginary, while $\frac{d}{dx}\zeta_{1+\eta}(x)$ has only even powers of $x$, and is therefore real when $x$ is imaginary. 

Using these results, it is straightforward to demonstrate that for an imaginary $x_0$ (the result of a tachyonic root existing in the $Q$-field KK tower), the right-hand side of Eq.(\ref{SSBLowestOrder}) consists of a real quantity divided by an imaginary quantity. Hence, to lowest order, we see the correction to a tachyonic root is purely imaginary. In the regime where Eq.(\ref{SSBLowestOrder}) represents a valid approximation of Eq.(\ref{SSBBoundary}) near $x_0$ then, it is therefore unrealistic to expect that a tachyonic root will be eliminated by SSB: Any tachyonic root should merely be shifted slightly (by an $O(v^2/M_{KK}^2)$ correction) along the imaginary axis.

While this conclusion suggests that tachyons cannot be eliminated by SSB in a large region of parameter space, some care must be taken before we can dismiss this possibility out of hand. Notably, Eq.(\ref{SSBLowestOrder}) predicts a small $O(v^2/M_{KK}^2)$ correction to the root equation only when $\zeta_{\eta_q}(x_0)-x_0 \tau_\pi \zeta_{1+\eta_q}(x_0) \neq 0$, or more accurately, when $\zeta_{\eta_q}(x_0)-x_0 \tau_\pi \zeta_{1+\eta_q}(x_0)$ is larger in magnitude than the $v^2/M_{KK}^2$ suppressed term in the denominator. However, a significant region in parameter space will not satisfy these conditions; if, \eg, $\eta_q \approx \eta_Q$, then $\zeta_{\eta_q}(x_0)-x_0 \tau_\pi \zeta_{1+\eta_q}(x_0) \approx 0$ whenever $\zeta_{\eta_Q}(x_0)-x_0 \tau_\pi \zeta_{1+\eta_Q}(x_0) \approx 0$. If the predicted shift in the root from Eq.(\ref{SSBLowestOrder}) is no longer a small correction, then the perturbative method employed is obviously invalid. To address the region of parameter space where this can occur, we need to extend our analysis to second order in the difference $(x'_0-x_0)$.

For simplicity, we shall explicitly display the second-order calculation of $(x'_0-x_0)$ in the scenario where terms proportional to $\zeta_{\eta_{q,Q}}(x_0)-x_0 \tau_\pi \zeta_{1+\eta_{q,Q}}(x_0)$ are very close to zero and may be safely ignored. Qualitatively, we expect that when this condition does not hold, the parameter space will rapidly approach the regime where Eq.(\ref{SSBLowestOrder}) is valid, which has already been addressed above. Expanding Eq.(\ref{SSBBoundary}) to second order in $(x'_0-x_0)$, then, yields a quadratic formula which can be solved for $(x'_0-x_0)$. The result yields two solutions for $(x'_0-x_0)$, which are given up to $O(v^2/M_{KK}^2)$ by
%
%
\begin{equation}\label{SSBSecondOrder} 
    (x'_0-x_0) \approx \left( \frac{v^2|Y|^2}{2 M_{KK}^2}\alpha \pm \frac{v |Y|}{\sqrt{2}M_{KK}}\sqrt{\beta \gamma}\right)\frac{1}{\gamma},
\end{equation}
where
\begin{align}
    \alpha &\equiv \frac{1}{2}\frac{d}{dx}(\zeta_{1+\eta_Q}(x)\zeta_{1+\eta_q}(x))|_{x=x_0}, \nonumber\\
    \beta &\equiv \sqrt{\zeta_{1+\eta_Q}(x_0) \zeta_{1+\eta_q}(x_0)},\\
    \gamma &\equiv \frac{d}{dx}(\zeta_{\eta_Q}(x)-x \tau_\pi \zeta_{1+\eta_Q}(x))|_{x=x_0}\frac{d}{dx}(\zeta_{\eta_q}(x)-x \tau_\pi \zeta_{1+\eta_q}(x))|_{x=x_0}. \nonumber
\end{align}
In the case of an imaginary $x_0$, $\alpha$ in Eq.(\ref{SSBSecondOrder}) takes on an imaginary value, while $\gamma$, as the product of two imaginary quantities, will be real. Thus, the first term on the right-hand side of Eq.(\ref{SSBSecondOrder}), $[v^2|Y|^2/(2M_{KK}^2)](\alpha/\gamma)$, is purely imaginary. The second term, $[v |Y|/(\sqrt{2} M_{KK})](\sqrt{\beta \gamma}/\gamma)$ has a complex phase governed by the term in the square root ($\gamma$, as noted before, is real). Because it is the product of four imaginary numbers, $\beta \gamma$ is necessarily real. However, whether or not this yields a real or imaginary correction is dependent on the sign of $\beta \gamma$. In practice, it appears that $\sqrt{\beta \gamma}$ is more likely to be real. For example, in the event that $\eta_Q \approx \eta_q$, we see that $\beta \gamma$ is the product of two squares of imaginary quantities. Since any imaginary number squared is negative, this implies that $\beta \gamma$ is the product of two negative numbers, and is therefore positive. However, we remind the reader that the second-order correction is still highly suppressed (in this case the real part of the correction is suppressed by $\sim v/M_{KK}$, while the imaginary correction is suppressed by $v^2/M_{KK}^2$), even when the approximation in Eq.(\ref{SSBLowestOrder}) breaks down. Given that the tachyonic roots we have found are generally of $O(1)$ (in units of $i$), this makes it exceedingly unlikely that any perturbative correction could convert a tachyonic root into a real root; it will either be shifted along the imaginary axis or slightly rotated into the complex plane.

\section{Summary}\label{Summary}

In this paper, we have closely examined the parameter space of the RS model with bulk fields for the unphysical regions which contain ghost and/or tachyon states. In general, we have found that the TeV-brane localized kinetic term, $\tau_\pi$, must be non-negative, \ie, $\tau_\pi \geq 0$, in order for the theory to be physical. By separating the problem into three distinct regions, we have then found further restrictions, summarized below (it should be noted that for highly TeV-brane localized fermions, \ie, the region where $\eta \lesssim -0.1$, the above restriction on $\tau_\pi$ is the only restriction to render the model physical).
For $-0.1 \lesssim \eta \lesssim 0.1$ (close to gauge-like localization)
\begin{equation}\label{CloseToGaugeT0}
\tau_0 > \frac{1}{2 \eta}\bigg[\bigg( \frac{\epsilon x_{max}}{2} \bigg)^{2 \eta} \frac{\Gamma(1-\eta)}{\Gamma(1+\eta)}-1 \bigg].
\end{equation}
Note that for $\eta=0$ (gauge bosons), this condition reduces to:
\begin{equation}
\tau_0 > \gamma+\log(\epsilon)+\log(\frac{x_{max}}{2}).
\end{equation}
For $\eta \gsim 0.1$ (highly UV-brane localized fermions)
\begin{equation}
\tau_0 > -\frac{1}{2 \eta}.
\end{equation}
Note that for $\eta=1$ (bulk gravitons), the conditions become
\begin{equation}
   \tau_\pi \leq 1
\end{equation}
and
\begin{equation}
    \tau_0 > -\frac{1}{2},
\end{equation}
where the upper bound $\tau_\pi \leq 1$ is required to avoid radion ghost states. Notably, the conditions for $\eta \approx 0$ will, as $\eta$ moves toward $-1$ or $1$, flow into the conditions for highly TeV-brane or UV-brane localized fermions, respectively. As a result, one can safely employ the conditions Eq.(\ref{CloseToGaugeT0}) and $\tau_\pi \geq 0$ as universal conditions for avoiding ghosts and tachyons, as long as $|\eta|<1$.

\begin{figure}[htbp]
\centerline{\includegraphics[width=4.5in]{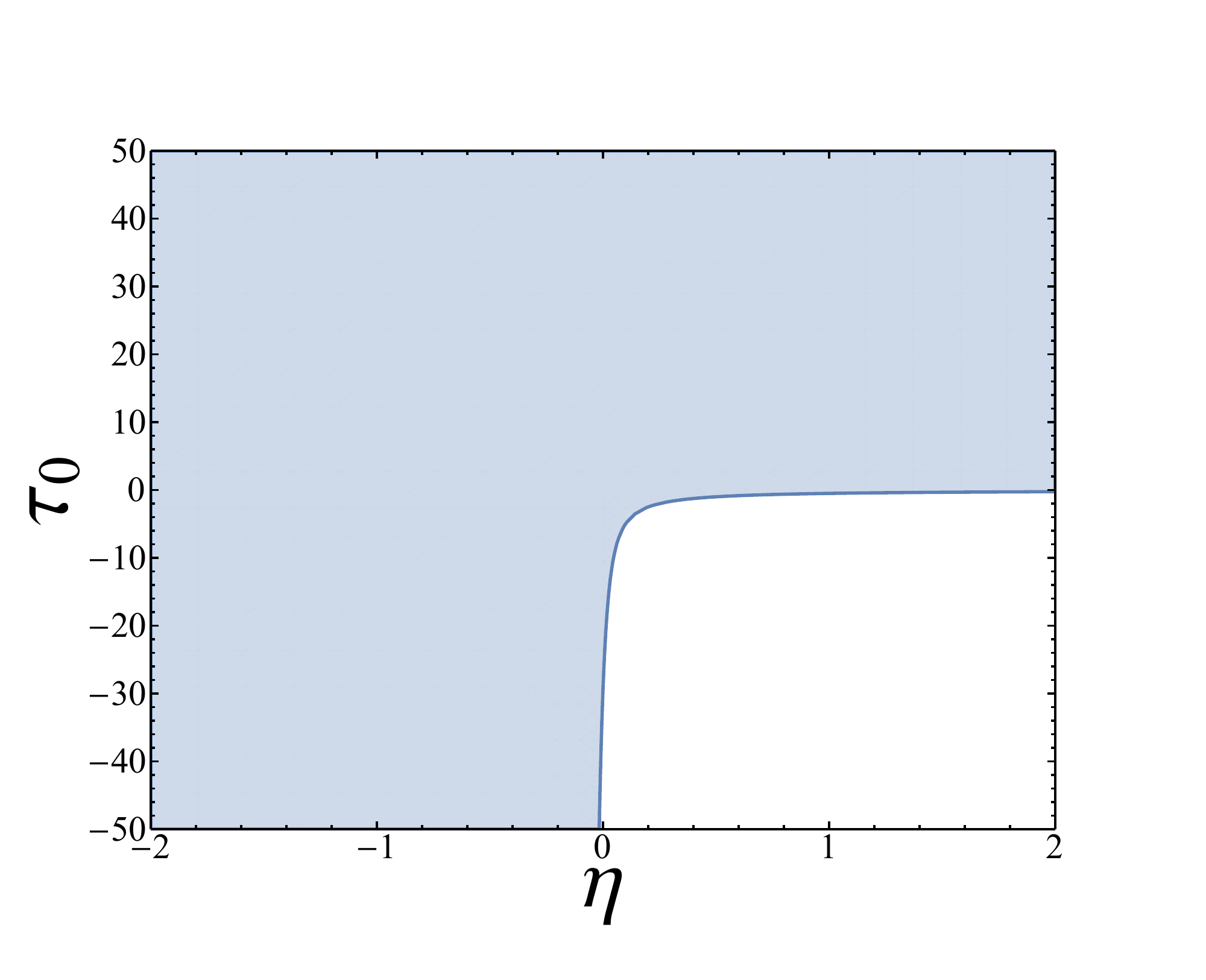}}
\vspace*{-0.10cm}
\caption{Region of parameter space which is free of both tachyonic modes and ghosts for $x_{max}=500$, $\tau_\pi \geq 0$. The blue region indicates physically allowable points in parameter space, while the unshaded region contains either tachyonic KK modes or ghost-like particles.}
\label{figTachyonSafe2}
\end{figure}

We combine our results in Fig. (\ref{figTachyonSafe2}) where we show the allowed parameter space of $\tau_0$ and $\eta$ for all $\tau_\pi \geq 0$, $|\tau_0|<50$, and $|\eta|<2$. Here, the shaded region in the Figure represents the physically allowable region of parameter space, assuming a cutoff of $x_{max}=500$. Note that the restrictions on $\tau_0$ depend only weakly on $x_{max}$, excluding a slightly larger region as $x_{max}$ increases, but since this dependence is so weak (it is only manifest near $\eta = 0$, and is proportional to $x_{max}^{2 \eta}$ for small $\eta$), other choices of $x_{max}$ result in qualitatively similar allowed regions. Furthermore, note that the universal restriction on $\tau_\pi$ is simply $\tau_\pi \geq 0$.

Finally, we have established that these constraints are reasonably robust against the introduction of SSB via the Higgs mechanism, indicating that these bounds also function as good approximations even when fermionic fields are granted mass via this mechanism. Notably, the introduction of SSB should introduce only small perturbations in the conditions to avoid ghost states and tachyons; in other words, the presence of SSB does not change the allowed regions of parameter space.

Overall, the restrictions on the RS parameter space derived above have far-reaching consequences for the future of RS model building. Notably, negative IR brane terms, featured in a number of analyses, \eg, \cite{Davoudiasl:2003zt,Davoudiasl:2002ua,hewett2016}, are entirely disallowed by the existence of tachyonic KK modes. Furthermore, the restrictions on the allowed parameter space for brane-localized kinetic terms, particularly in the gauge boson sector, limit their ability to ameliorate constraints on models arising from precison electroweak measurements, lending credence to the necessity for a bulk custodial symmetry (such as that discussed in \cite{Casagrande:2010}) to preserve these quantities in models with bulk SM fields in the warped extra dimension.

$\textbf{Acknowledgements: }$

We would like to thank Adam Falkowski for pointing out the relevance of this subject. This work was supported by the Department of Energy, Contract DE-AC02-76SF00515.



\begin{thebibliography}{99}

\bibitem{Randall:1999ee} 
  L.~Randall and R.~Sundrum,
  Phys.\ Rev.\ Lett.\  {\bf 83}, 3370 (1999)
  [hep-ph/9905221].

\bibitem{eDims}
  For a review of extra dimensional models, see, J.~L.~Hewett and M.~Spiropulu,
  Ann.\ Rev.\ Nucl.\ Part.\ Sci.\ {\bf 52}, 397 (2002)
  [hep-ph/0205106];
  C.~Csaki, 
  In *Shifman, M.~(ed.) et al.: From fields to strings, vol.~2* 967-1060
  [hep-ph/0404096];
  E.~Ponton, arXiv:1207.3827 [hep-ph].

\bibitem{huberHierarchy}
  S.~J.~Huber and Q.~Shafi,
  Phys.\ Lett.\ B {\bf 498} 256-262 (2001)
  [hep-ph/0010195].

\bibitem{gherghetta}
  T.~Gherghetta and A.~Pomarol,
  Nucl.\ Phys.\ B {\bf 586}, 141 (2000)
  [hep-ph/0003129].

\bibitem{casagrande}
  S. Casagrande, F. Goertz, U. Haisch, M. Neubert, and T. Pfoh,
  JHEP {\bf 10} (2008) 094
  [hep-ph/0807.4937].
  
\bibitem{huber}
  S.~J.~Huber,
  Nucl.\ Phys.\ B {\bf 666} 269-288 (2003)
  [hep-ph/0303183].
  
\bibitem{Davoudiasl:2000wi} 
  H.~Davoudiasl, J.~L.~Hewett and T.~G.~Rizzo,
  Phys.\ Rev.\ D {\bf 63}, 075004 (2001)
  [hep-ph/0006041].
  
\bibitem{carenaEW}
  M.~Carena, E.~Ponton, J.~Santiago, and C.~E.~M.~Wagner,
  Phys.\ Rev.\ D {\bf 76}, 035006 (2007)
  [hep-ph/0701055].
  
\bibitem{agashe}
  K.~Agashe, A.~Delgado, M.~J.~May and R.~Sundrum,
  JHEP {\bf 0308}, 050 (2003)
  [hep-ph/0308036].
  
\bibitem{hewett}
  J.~L.~Hewett, F.~J.~Petriello, T.~G.~Rizzo,
  JHEP {\bf 0209} 030 (2002)
  [hep-ph/0203091].
  
\bibitem{rizzoEW}
  T.~G.~Rizzo and J.~D.~Wells, 
  Phys.\ Rev.\ D {\bf 61}, 016007 (2000)
  [hep-ph/9906234].
  
\bibitem{Dey}
  U.~K.~Dey and T.~S.~Ray,
  Phys.\ Rev.\ D {\bf 93}, 011901 (2016)
  arXiv:1507.04357 [hep-ph].
  
\bibitem{Casagrande:2010}
  S.~Casagrande, F.~Goertz, U.~Haisch, M.~Neubert, and T.~Pfoh,
  JHEP {\bf 1009} 014 (2010)
  arXiv:1005.4315 [hep-ph].

\bibitem{Grossman:2000}
  Y.~Grossman and M.~Neubert,
  Phys.\ Lett. \ B {\bf 474}, 361 (2000)
  [hep-ph/9912408].

\bibitem{georgi}
  H.~Georgi, A.~K.~Grant, and G.~Hailu,
  Phys.\ Lett.\ B {\bf 506}, 207 (2001)
  [hep-ph/0012379].
  
\bibitem{carenaBranes}
  M.~Carena, E.~Ponton, T.~Tait, C.~E.~M.~Wagner,
  Phys.\ Rev.\ D {\bf 67}, 096006 (2003)
  [hep-ph/0212279].
  
\bibitem{Aguila:2003}
  F.~del~Aguila, M.~Perez-Victoria and J.~Santiago,
  JHEP 0302 {\bf 051} (2003)
  [hep-th/0302023].
  
\bibitem{Davoudiasl:2003zt} 
  H.~Davoudiasl, J.~L.~Hewett and T.~G.~Rizzo,
  JHEP {\bf 0308}, 034 (2003)
  [hep-ph/0305086].
  
\bibitem{Davoudiasl:2002ua} 
  H.~Davoudiasl, J.~L.~Hewett and T.~G.~Rizzo,
  Phys.\ Rev.\ D {\bf 68}, 045002 (2003)
  [hep-ph/0212279].

\bibitem{Goldberger}
  W.~D.~Goldberger and M.~B.~Wise,
  Phys.\ Rev.\ Lett.\ {\bf 83}, 4922 (1999)
  [hep-ph/9907447].  
  
\bibitem{Carena:2004zn} 
  M.~Carena, A.~Delgado, E.~Ponton, T.~M.~P.~Tait and C.~E.~M.~Wagner,
  Phys.\ Rev.\ D {\bf 71}, 015010 (2005)
  [hep-ph/0410344],
  F.~del Alguila, M.~Perez-Victoria, and J.~Santiago,
  [hep-ph/0305119].
  
\bibitem{dillon}
  B.~M.~Dillon, D.~P.~George, and K.~L.~McDonald,
  Phys.\ Rev.\ D {\bf 94} 064045 (2016)
  arxiv:1605.03087 [hep-ph].
  
\bibitem{rizzoHD}
  T.~G.~Rizzo,
  JHEP {\bf 0501} 028 (2005)
  [hep-ph/0412087].
  
\bibitem{mathematica}
  Wolfram Research, Inc.,
  Mathematica, Version 10.2,
  Champaign, IL (2015).
  
\bibitem{bulkgauges}
  H.~Davoudiasl, J.~L.~Hewett and T.~G.~Rizzo,
  Phys.\ Lett.\ B {\bf 473}, 43 (2000)
  [hep-ph/9911262];
  A.~Pomarol,
  Phys.\ Lett.\ B {\bf 486}, 153 (2000)
  [hep-ph/9911294].
  
\bibitem{hewett2016}
  J.~L.~Hewett and T.~G.~Rizzo,
  arXiv:1603.08250 [hep-ph].
  


  
  


  

  

  

  


\end{thebibliography}
\end{document}